\documentclass[journal]{IEEEtran}

\ifCLASSINFOpdf
\usepackage[pdftex]{graphicx}

\else

\fi

\usepackage{float}
\usepackage[cmex10]{amsmath}
\usepackage{amssymb}   % \triangleq vs. gibi seyler bunda
\usepackage{amsxtra}
\usepackage{amscd}
\usepackage{amsthm}
 \usepackage{amsxtra}
 \usepackage{amscd}
 \usepackage{steinmetz}
 \usepackage{amsthm}
\usepackage{tikz}
\usetikzlibrary{matrix}
\usepackage{stfloats}
\usepackage{graphicx, subfigure}   % sekilller icin
\usepackage{relsize}
\usepackage{algorithm,algpseudocode}

\usepackage{subfig}

\usepackage{array}  % tablo icin satir acma
\usepackage{authblk}
\usepackage{multirow}  % tablolar icin coklu satir kullanmada
\usepackage{arydshln}

\newtheorem{lemma}{Lemma}

\usepackage{color}

\usepackage{cite}
\renewcommand{\thefootnote}{\arabic{footnote}}

\usepackage{balance}

  \author{
  Ali Tugberk Dogukan,~\IEEEmembership{Graduate Student Member,~IEEE,} Emre Arslan,~\IEEEmembership{Graduate Student Member,~IEEE,} and Ertugrul Basar,~\IEEEmembership{Fellow,~IEEE}
\vspace{-0.6cm}  
  
  \thanks{The authors are with the Communications Research and Innovation Laboratory (CoreLab),  Department of Electrical and Electronics Engineering, Ko\c{c} University, Sariyer 34450, Istanbul, Turkey. Ali Tugberk Dogukan and Emre Arslan are also with ULAK Communications and Turkcell Iletisim Hizmetleri Inc., respectively. This work was also supported by The Scientific and Technological Research Council of Turkey (TUBITAK) through the 1515 Frontier Research and Development Laboratories Support Program under Project 5229901 - 6GEN. Lab: 6G and Artificial Intelligence Laboratory, and also Grant 120E401. In addition it has been funded by, ULAK Communications Inc. \mbox{Email: \{tugberk.dogukan@ulakhaberlesme.com.tr}, {emre.arslan@turkcell.com.tr}, {and ebasar@ku.edu.tr}\}
  } 
  
}

\begin{document}

\bstctlcite{IEEEexample:BSTcontrol}

	\title{{ Reconfigurable Intelligent Surface-Enabled Downlink NOMA  }}

	\maketitle	
	
	\begin{abstract}
Reconfigurable intelligent surfaces (RISs) bring great potential to the advancement of 6G and beyond wireless communication technologies. RISs introduce a great degree of flexibility, allowing some sort of virtual control over the wireless channel. Exploiting the flexibility introduced by RISs, we propose a novel RIS-enabled downlink (DL) non-orthogonal multiple access (NOMA) scheme where NOMA is enabled over-the-air rather than at the base station (BS) or the receiver (Rx). Here, the RIS is partitioned into distinctive groups where each part of the RIS serves a different user equipment (UE) to perform multiple accessing. The BS transmits an unmodulated signal to the RIS, and each partition modulates the impinging signal over-the-air by introducing a phase shift according to the incoming information bits to serve the corresponding UE. First, the end-to-end system model for the proposed system is presented. Furthermore, outage probability calculations, theoretical error probability analysis, and bit error rate (BER) derivations are discussed and reinforced with comprehensive computer simulation results.

	\end{abstract}

	\begin{IEEEkeywords}
	
		  6G, reconfigurable intelligent surface, NOMA, downlink, outage probability, BER, partitioning. 
		 
	\end{IEEEkeywords}

	\IEEEpeerreviewmaketitle
		\IEEEpubidadjcol
		
	\renewcommand{\thefootnote}{\fnsymbol{footnote}}

	\section{Introduction}

	\IEEEPARstart
{G}{lobally}, wireless communication technologies are in demand, and countries are striving to meet the strict expectations of 6G and beyond applications \cite{saad2019vision}. 
The advent of 6G technology is anticipated to facilitate the deployment of advanced applications, including but not limited to unmanned aerial vehicle (UAV) networks, intelligent environments, autonomous systems, collaborative sensing and communication systems, eHealth solutions, virtual reality experiences, and a myriad of other innovative applications \cite{de2021survey}. In addition, many applications are envisioned to co-exist rather than function independently \cite{al2023resource}. However, these applications require extremely high performance; for example, in autonomous systems, there is almost no room for error where reliability and latency are crucial since people's lives are at stake \cite{dang2020should}. To meet these stringent requirements, innovative solutions are proposed to enhance the performance of wireless communication systems; nonetheless, it must be noted that one proposed solution alone cannot be sufficient to enhance wireless communication performance as needed. Hence, the co-existence and integration of multiple innovative technologies must be considered. 

Reconfigurable intelligent surfaces (RISs) have recently emerged as a preeminent technology that has captivated the interest of both academia and industry \cite{basar2019wireless}. RISs are large surfaces with many elements that are made up of reflective meta-materials. These elements can contain PIN diodes, and when a voltage is applied, it allows the RIS to manipulate the electrical length and, hence, the phase of the impinging signal. Due to having intelligent control over its elements, one of the most significant advantages of RIS technology is its flexible structure \cite{elmossallamy2020reconfigurable}. It can easily be incorporated and integrated with existing wireless communication techniques such as index modulation \cite{basar2017index}, non-orthogonal multiple access (NOMA) \cite{dai2018survey}, different waveforms and much more. Furthermore, it is flexible in the modes it can operate in, which broadens the possible use cases of an RIS and makes it a desirable tool. An RIS can operate in reflect, refract, and absorb modes to the impinging signals for diverse intentions, allowing some degree of control over the channel with its reconfigurable elements \cite{tapio2021survey}. Furthermore, an RIS can operate in a hybrid manner and even be partitioned into multiple groups, with each group serving a different purpose. 

NOMA is another technology that has great potential for future wireless communication systems. This approach has been contemplated as a subject of investigation within the context of 5G; nevertheless, empirical validation is required to establish the capacity of NOMA to yield substantial enhancements in the realm of post-5G communications \cite{vaezi2019non}. NOMA allows multiple user equipments (UEs) to share valuable and scarce communication resources such as time, frequency, etc., by differentiating them in different domains, most popularly in code and power domains (PD) \cite{liu2021sparse,ek}. For example, through superposition coding and successive interference cancellation (SIC) methods, UEs can enjoy the benefits of NOMA in the power domain. NOMA also possesses flexibility in its applications and is an important tool that can be integrated into conventional communication systems such as index modulation (IM), orthogonal frequency division multiplexing (OFDM), RIS, and more \cite{arslan2020index}.

\subsection{Related Work}
\subsubsection{Studies on RIS}
RISs have been used as a tool to enhance the performance of various methods in wireless communication systems. In \cite{9201413}, the RIS orientation and horizontal distance have been investigated and optimized to maximize cell coverage. RISs are able to provide energy efficiency and spectrum efficiency, but there exists a trade-off based on UE preference. The non-trivial trade-off between these for multiple-input multiple output-uplink (MIMO-UL) communication is studied in \cite{9309152}. While conventionally, the equalization process is performed at the transmitter (Tx) or receiver (Rx), \cite{arslan2022over} shows that the RIS can even virtually equalize a signal in a frequency-selective scenario. An RIS can also be partitioned into different sections and groups where each partition can serve a different purpose. For a multi-UE scenario, the RIS elements are partitioned and allocated in an optimal manner to maximize UE fairness in \cite{khaleel2021novel}. On the other hand, an interesting study partitions the RIS for physical layer security in \cite{arzykulov2023artificial}. While one part of the RIS serves to enhance communication at the legitimate UE, the other part of the RIS introduces artificial noise for the illegitimate UE. In addition, practical studies have been conducted for RIS-aided systems such as in \cite{arslan2023networkindependent, kayraklik2022indoor}, where the RIS enables coverage and efficient RIS codebook generation algorithms are proposed. Lastly, a comprehensive survey that summarizes the designs of the machine and deep learning-assisted RIS communication systems is presented in \cite{ozpoyraz2022deep}.

\subsubsection{Studies on NOMA}

NOMA is a relatively older concept that has been and is still being investigated due to its large potential to enable multiple UEs to share valuable communication resources. Optimal power allocation for UEs is critical in NOMA and the authors in \cite{timotheou2015fairness} and \cite{yang2016general} address this issue by investigating power allocation and fairness for UEs. This is also considered for UE clusters in \cite{ali2016dynamic}, which proposes a low-complexity and sub-optimal grouping scheme and then deriving a power allocation policy to maximize the overall system throughput considering both UL and downlink (DL). Furthermore, comprehensive error probability analyses with channel errors, bit-error-rate (BER) performance investigations with SIC errors, and other detailed theoretical derivations have been conducted, giving NOMA studies a solid theoretical reference \cite{ferdi2020error, kara2018ber, kara2019performance, kara2020improved}. The literature regarding NOMA is very rich  due to its flexibility in means of coexisting with conventional technologies such as IM, machine learning, and even RIS \cite{arslan2020index, arslan2022reconfigurable, gui2018deep}.

\subsubsection{Studies on RIS$+$NOMA}

As stated previously, RIS and NOMA are both considerably flexible technologies that can be integrated with current wireless communication systems. The main motivation behind amalgamating RIS and NOMA is to fully embrace both flexible technologies potential and supplement each other while minimizing specific drawbacks either technology may have individually. For example, in conventional power domain NOMA systems, superposition coding and SIC is required to remove interference of other signals. However, the SIC process comes with unavoidable SIC errors which significantly degrade the practicality and performance of the system. RIS technology may provide alternative solutions to combat this issue. In addition, RIS technology is severely impacted by double fading path loss, hardware complexities, and other issues that the flexibility of NOMA may be able to alleviate. These two technologies have been further improved and merged to provide significant gains and new perspectives. For example, while the power difference required in power domain NOMA is generally conducted at the Tx or Rx, a hybrid RIS with active and passive partitions has been utilized to enable the power difference over the air in UL scenario \cite{arslan2022reconfigurable}. In \cite{chauhan2022ris} and \cite{kumar2023ris}, two RIS partitioned systems are introduced to maximize diversity order and improve signal quality, and an RIS assisted UE pairing THz-NOMA system has been proposed that pairs UEs based on the distance and power allocation between them to enhance their bit error rate (BER) performance, respectively. Other works also investigate the physical layer security for RIS-NOMA systems, joint active and passive beamforming for MISO RIS-aided NOMA to maximize the sum rate \cite{mu2020exploiting},  joint location and beamforming designs for STAR-RIS assisted NOMA systems, and attach an RIS to an unmanned aerial vehicles (UAV) to assist in communication to other base station (BS) to UAV or UE communications \cite{song2021physical, gao2023joint, bariah2023performance}. In addition, an effective capacity analysis regarding an RIS and NOMA network is shown in \cite{li2021effective}. In \cite{aldababsa2023simultaneous}, simultaneous transmitting and reflecting RIS (STAR-RIS)-aided NOMA scheme has been proposed.

\subsection{Main Contributions}

In this paper, we propose a DL over-the-air NOMA scheme with the aid of an RIS to serve multiple UEs. It should be noted that in conventional DL PD-NOMA, the power difference is enabled at the BS through superposition coding. However, for the first time, DL-NOMA is enabled at the RIS using over-the-air modulation concept in the proposed scheme. To emphasize, the power difference required to enable power domain NOMA is not generated at the BS but rather we generate this power difference over-the-air with the help of the RIS as well as apply modulation to the impinging unmodulated signal. Here, the RIS elements are equally divided into multiple groups for enabling multiple-accessing where the number of groups is equal to the number of UEs. Note that each group serves one UE. An unmodulated carrier signal is transmitted from the BS to the RIS. Each partition of the RIS coherently aligns the phases of the impinging signal and modulates it using the information bits received by the BS through the backhaul link. Hence, a power level difference is generated at each UE since signals reflected by the elements of each RIS partition are constructively summed up. Even though there is interference of the reflected signals corresponding to the other UEs, they become negligible in comparison to the coherently aligned signal for the specified UE; and this signal can be directly decoded. In contrast to conventional NOMA schemes, the proposed scheme does not require SIC to remove the interference of the other UEs, hence, reducing the complexity and improving the efficiency of the system. It should be noted that SIC is a significant issue in conventional NOMA that dramatically increases the complexity and reduces the performance of wireless communications systems where the drawbacks exponentially increase with the increase of the number of UEs. In contrast to classical PD-NOMA, the proposed scheme directly decodes the signal without applying SIC and its decoding complexity does not increase with the number of UEs.

The major contributions of this paper are listed as follows:
\begin{itemize}
        \item A novel communication system is proposed to enable DL NOMA over-the-air with a single passive RIS.
	    \item A thorough end-to-end system model of the proposed scheme is provided and analyzed along with computer simulations.
      \item The outage probability is derived and calculated for different configurations and confirmed with simulations. 
	    \item A theoretical error probability analysis is provided along with computer simulation results for varying parameters and scenarios.
	    \item Computer simulations have been performed with perfect and imperfect channel state information (CSI) conditions to assess the BER performance.
	    \item Sum-rate has been presented for the proposed NOMA scheme.
        \item Comparisons for all of the above analysis have been conducted against conventional passive RIS orthogonal multiple access (OMA) scenarios existing in the literature.
     \end{itemize}

\subsection{Paper Organization}
The rest of the paper is organized as follows, Section II presents the end-to-end system model of the proposed scheme. Section III presents a theoretical analysis of the outage probability and theoretical BER performance. Furthermore, comprehensive computer simulation results investigating outage probability, sum rate, and BER performances are provided and discussed in Section IV. Finally, the paper is concluded with the final remarks in Section V.

$\emph{Notations}$: Bold, lowercase, and capital letters stand for vectors and matrices, respectively. $\left(\cdot\right)^{\mathrm{T}}$ is used for transposition operation. $X \sim \mathcal{CN}\left(\mu_X,\sigma_{X}^2\right)$ denotes the distribution of a circularly symmetric complex
Gaussian random variable $X$ with mean $\mu_X$ and variance $\sigma_{X}^2$. $\lVert \cdot \rVert$ and $E [\cdot]$ are used for Euclidean norm and expectation, respectively. $\lvert \mathbb{X} \rvert$ is the cardinality of set $\mathbb{X}$. $f_X\left(x\right)$, $F_X\left(x\right)$, $\Psi_X\left(\omega\right)$, and $M_X\left(s\right)$ respectively represent probability density function (PDF), cumulative distribution function (CDF), characteristic function (CF), and moment generating function (MGF) of a random variable $X$. $\mathrm{Pr}\left[\cdot\right]$ denotes the probability of an event.

\begin{figure}[t]
		\centering
		\includegraphics[width=0.95 \columnwidth]{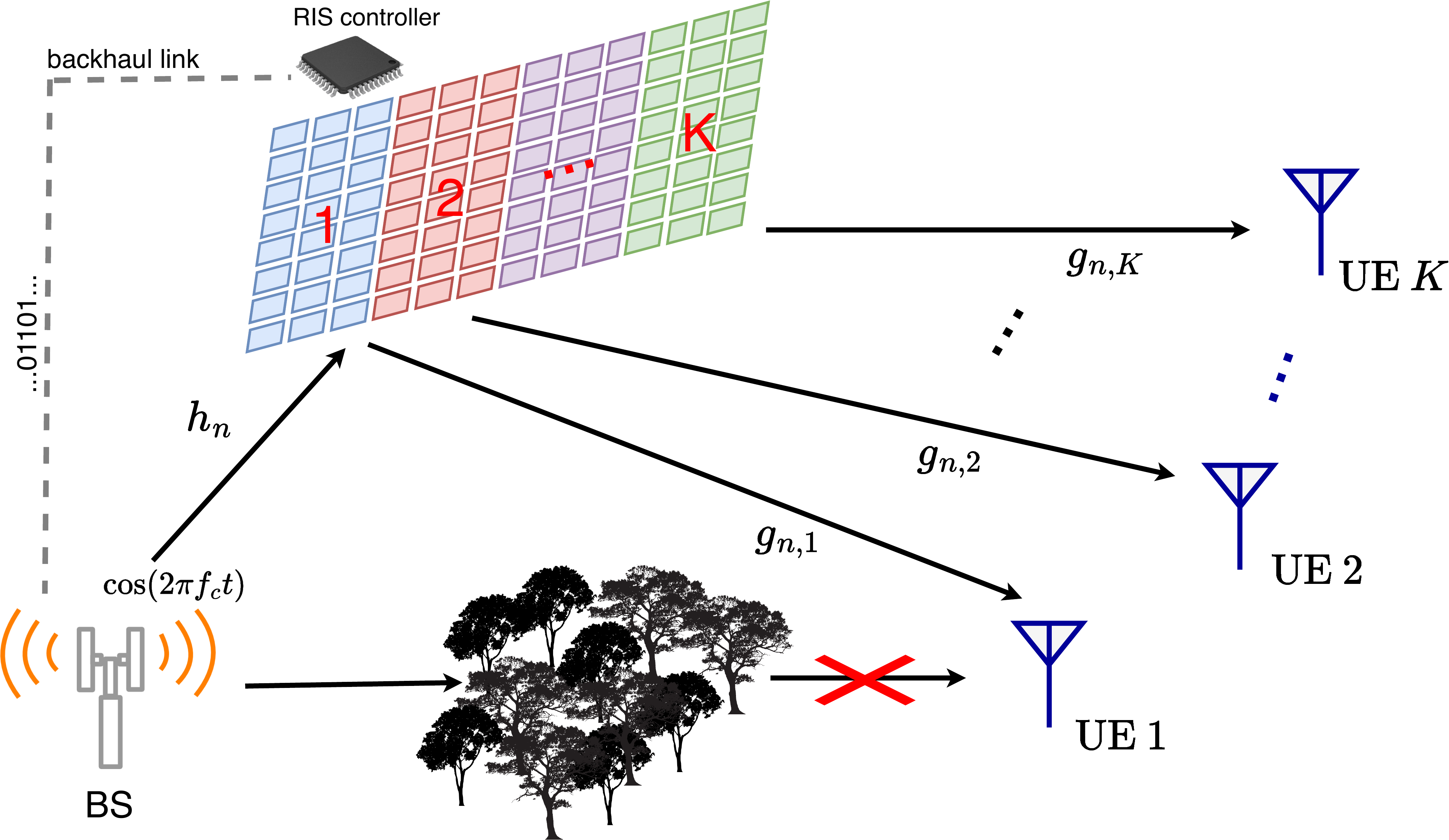} 
            \vspace{-0.2cm}
		\caption{ The system model of the proposed DL NOMA scheme that serves $K$ UEs.}
		\label{fig:Fig1}
\end{figure}

\section{ System Model}

%\begin{figure*}[t]
%    \centering
%    \includegraphics[width=\linewidth]{SystemNew.drawio.pdf}
%    \vspace{-0.2cm}
%    \caption{Transceiver architecture of CI-OFDM-RIQIM.}
%    \label{fig:system_model}
%\end{figure*}

\begin{figure*}[bh!]

\begin{align} \tag{2}
    y_k = \sqrt{P_t} \Bigg [ \underbrace{e^{j \xi_k}\sum_{n \in \mathbb{I}_k} \alpha_n \beta_{n,k}}_{\textrm{Useful term}}   +  \underbrace{ \sum_{k' \neq k}
 \sum_{n\in \bar{\mathbb{I}}_{k}} \alpha_n \beta_{n,k} e^{\left(j \xi_{k'} - \psi_{n,k'} + \psi_{n,k}\right)} }_{\textrm{Interference of other RIS parts}}
    \Bigg ] + w_k.
    \label{eq2}
\end{align}
\end{figure*}

As seen in Fig. 1, a DL NOMA system is considered, where the BS and an RIS simultaneously serve $K$ UEs under frequency-flat Rayleigh fading channels. Therefore, it is assumed that the LoS link between the BS and UEs is blocked via obstacles such as buildings or trees, and a backhaul link exists between the BS and RIS. As in \cite{araghi2022reconfigurable}, the RIS can be used for coverage extension when LoS does not exist between the BS and UEs. Hence, this assumption is valid.  To serve $K$ UEs, the RIS with $N$ elements is separated into $K$ partitions, each including $N_g$ number of elements, i.e., $N_g=N/K$. The BS transmits an unmodulated signal with a transmit power ($P_t$) and a carrier frequency ($f_c$). The $k^{\mathrm{th}}$ RIS partition modulates the incoming unmodulated signal according to the information bits of the $k^{\mathrm{th}}$ UE conveyed through the backhaul link between the BS and RIS by introducing an additional phase shift, $k=1,\cdots,K$. At the $k^{\mathrm{th}}$ partition of the RIS, a total number of $\log_2\left(M_k\right)$ incoming bits introduce a phase shift $\xi_k$ for the $k^{\mathrm{th}}$ UE, where $M_k$ and $\xi_k$ are the modulation order and phase value of the determined symbol drawn from $M_k$-PSK constellation for the $k^{\mathrm{th}}$ UE, respectively. Hence, the RIS can be considered as a part of the BS which enables a low-cost transmitter architecture as in\cite{basar2019wireless, basar2019transmission}. The received signal at the $k^{\mathrm{th}}$ UE can be obtained as:
\begin{equation}
    y_k = \sqrt{P_t} \sum_{n=1}^{N} g_{n,k} e^{j \theta_n} h_n + w_k,
    \label{eq1}
    \vspace{-0.2cm}
\end{equation}
where $h_n = \alpha_n e^{j \phi_n}$ and $g_{n,k} = \beta_{n,k} e^{j \psi_{n,k}}$ are the independent and identically distributed (i.i.d.) Rayleigh fading channel coefficients between the BS and $n^{\mathrm{th}}$ RIS element and between the $n^{\mathrm{th}}$ RIS element and $k^{\mathrm{th}}$ UE, respectively, $n=1,\cdots,N$. Moreover, $w_k$ and $\theta_n$ stand for the additive white Gaussian noise sample, which follows the distributions $\mathcal{CN}\left(0, N_0\right)$ and adjustable phase introduced by the $n^{\mathrm{th}}$ RIS element, respectively. Since the $k^{\mathrm{th}}$ RIS partition serves the $k^{\mathrm{th}}$ UE, ${\bar{n}}^{\mathrm{th}}$ RIS element is coherently aligned and modulated as $\theta_{\bar{n}} = \xi_k - \phi_{\bar{n}} - \psi_{{\bar{n}},k}$, ${\bar{n}} \in \mathbb{I}_k$, where $\mathbb{I}_k = \left\{ N_g\left(k-1\right)+1, N_g\left(k-1\right)+2, \cdots, N_g k  \right\}$, is a set of indices of the RIS elements allocated to serve the $k^{\mathrm{th}}$ UE, $k=1,\cdots,K$, $	\lvert \mathbb{I}_k \rvert = N_g$. Note that the distributions of $h_n$ and $g_{n,k}$ follow $\mathcal{CN}\left(0,\sigma^2_h\right)$ and $\mathcal{CN}\left(0,\sigma^2_{g_k}\right)$, respectively. Assuming $\mathbb{I}_{k'} = \mathbb{I} - \mathbb{I}_{k}$, $\mathbb{I}=\left\{1,2,\cdots,N\right\}$, $\lvert \mathbb{I}_{k'} \rvert = \left(K-1\right)N_g$, after determining the phase values of the RIS elements, (\ref{eq1}) can be reexpressed as in (\ref{eq2}). As we may know, the sum of real numbers increases monotonically, on the other hand, the sum of complex numbers is not monotonic due to phase terms resulting in fluctuations in their sum. The useful term in (\ref{eq2}) is the sum of real numbers instead of complex numbers as in interference term.  Hence, as mentioned previously, the useful term monotonically increases while the interference term does not grow and becomes negligible for a large value of $\left\lvert \mathbb{I}_k \right\rvert$.

At the $k^{\mathrm{th}}$ UE, the maximum likelihood (ML) detection rule is used to decode $y_k$ as below:
\setcounter{equation}{2}
\begin{equation} 
    \label{eq3}
    \hat{\mu}_k = \arg \underset{\mu=1,\cdots,M_k}{\min} \left\| y_k - e^{j \xi_k(\mu)} \Delta \right\|^2,
    \vspace{-0.2cm}
\end{equation}
where $\Delta = \sqrt{P_t} \sum_{n \in \mathbb{I}_k} \alpha_n \beta_{n,k} $ and $\xi_k(\mu)$ is the phase value of the $\mu^{\mathrm{th}}$ symbol drawn from $M_k$-ary PSK constellation. Finally, $\hat{\mu}_k$ is demapped into $\log_2\left(M_k\right)$ information bits. Note that SIC is not required to decode the received signal in the proposed scheme; however, it is a vital step in the power domain NOMA systems \cite{ding2020unveiling}.

The SINR of the $k^{\mathrm{th}}$ UE can be derived from (\ref{eq2}) as follows: 
\begin{equation} 
 \label{eq4}
    \gamma_k = \frac{P_t \left| \sum_{n\in \mathbb{I}_k}  \alpha_n \beta_{n,k}  \right|^2}{P_t  \left | \sum_{k' \neq k}
 \sum_{n\in \mathbb{I}_{k'}} \alpha_n \beta_{n,k} e^{(j \xi_{k'} - \psi_{n,k'} + \psi_{n,k})}  \right|^2 + N_0}.
\end{equation}

It should be noted that when there exists a direct path between the BS and UE, the received signal at the $k^{\mathrm{th}}$ UE can be given as:
\begin{align}
    y_k = \sqrt{P_t} \left(f_k + \sum_{n=1}^{N} g_{n,k} e^{j \theta_n} h_n \right) + w_k,
    \label{eq1}
\end{align}
where $f_k$ is the Rayleigh fading coefficient between the BS and $k^{\mathrm{th}}$ UE. In this case, at the $k^{\mathrm{th}}$ UE, ML rule as in (\ref{eq3}) can be used as follows:
\begin{equation} 
    \label{eq3}
    \hat{\mu}_k = \arg \underset{\mu=1,\cdots,M_k}{\min} \left\| y_k - \sqrt{P_t} f_k - e^{j \xi_k(\mu)} \Delta \right\|^2,
\end{equation}
Here, the unmodulated signal term $\sqrt{P_t} f_k $ arriving directly from the BS can easily be extracted to remove the interference effect of LoS path. 

Additionally, when compared to NOMA clustering studies where RIS partitions serve multiple users in a cluster rather than one user, in general, the proposed scheme is expected to provide enhanced performance since each group of the RIS focuses on serving a single UE. Whereas in a clustering approach, the RIS needs to optimize itself for numerous UEs at the same time in a cluster, which will inevitably come with significant trade-offs.

\begin{figure}[t]
    \centering
    \includegraphics[width=0.8\columnwidth]{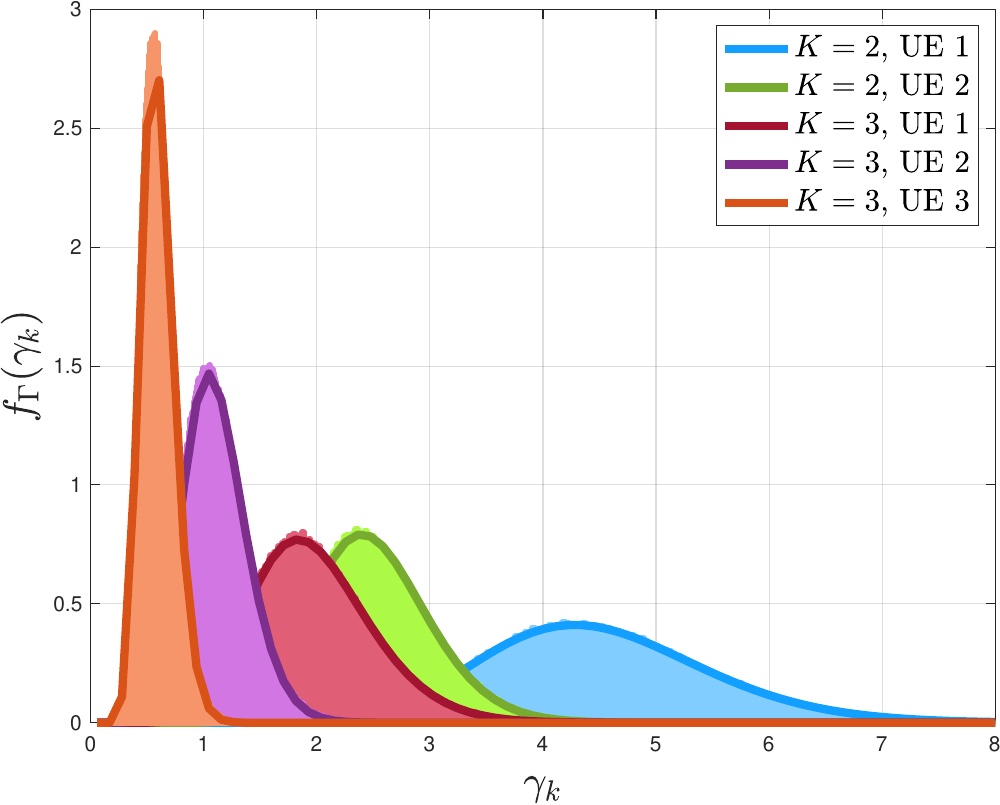}
    \vspace{-0.2cm}
    \caption{Distribution fit of SINR for varying UEs. \vspace{-0.2cm}}
    \label{fig:distribution}
\end{figure}

\section{Performance Analysis}

In this section, we provide a comprehensive theoretical outage probability and BER analysis for the proposed scheme.  Using the SINR, we approach the outage probability analysis through the use of Gil-Pelaez's inversion formula to obtain the CDF for each user. The SINR distribution is confirmed theoretically along with the distribution fitter tool on MATLAB. Using the parameters of the fitted distribution, we obtain the BER analysis.    

\subsection{Theoretical Outage Probability Analysis}

This subsection analyzes the theoretical outage probability of the RIS-DL-NOMA scheme. For convenience, the SINR of the $k^{\mathrm{th}}$ UE can be rewritten as:
\begin{align}
\gamma_{k} &= \frac{P_t \left| \mathcal{A} \right|^2 }{P_t \left| \mathcal{B} \right|^2 + N_0}, \\
\mathcal{A} &=  \sum_{n\in \mathbb{I}_k}  \alpha_n \beta_{n,k},  \\
\mathcal{B} &=  \sum_{k' \neq k}
 \sum_{n\in \mathbb{I}_{k'}} \delta_{n,k,k'}, 
\end{align}
where $\delta_{n,k,k'} = \alpha_n \beta_{n,k} e^{(j \xi_{k'} - \psi_{n,k'} + \psi_{n,k})}$ and
the amplitudes of channel coefficients of $h_n$ and $g_{n,k}$ ($\alpha_n$ and $\beta_{n,k}$) are assumed to be independently Rayleigh distributed random variables, and $\mathrm{E}[\alpha_n \beta_{n,k}] = \sigma_h \sigma_{g_k} \frac{\pi}{4}$ and $\mathrm{VAR}[ \alpha_n \beta_{n,k} ] = \sigma^2_h \sigma^2_{g_k} \left(1-\frac{\pi^2}{16}\right)$. For a sufficiently large value of $N_g$, $\mathcal{A}$ follows Gaussian distribution with $\mathcal{A} \sim \mathcal{N} \left(N_g \sigma_h \sigma_{g_k} \frac{\pi}{4}, N_g \sigma^2_h \sigma^2_{g_k} \left(1-\frac{\pi^2}{16}\right) \right)$, due to the central limit theorem (CLT). Moreover, $|\mathcal{A}|^2$ becomes a non-central chi-square random variable with one degree of freedom. In contrast, since complex terms exist in 
$\delta_{n,k,k'}$, $\mathcal{B}$ converges to complex Gaussian distribution with $\mathcal{B} \sim  \mathcal{CN} \big(0, N_g (K-1) \sigma^2_h \sigma^2_{g_k}  \big)$ and $|\mathcal{B}|^2$ becomes a central chi-square random variable with two degrees of freedom. Therefore, the SINR of the $k^{\mathrm{th}}$ UE $(\gamma_k)$ is the ratio of two chi-square random variables. 

\begin{lemma}
		\label{lemma:OPA}
		The mean and variance values of $\mathcal{A}$ can be given as $\mu_\mathcal{A}=N_g \sigma_h \sigma_{g_k} \frac{\pi}{4}$ and $\sigma_\mathcal{A}=N_g \sigma^2_h \sigma^2_{g_k} \left(1-\frac{\pi^2}{16} \right)$, respectively. 
	\end{lemma}\begin{proof}
		Readers may refer to Appendix A for the derivation steps.
	\end{proof} 
	
\begin{lemma}
		\label{lemma:OPB}
		The mean and variance values of $\mathcal{B}$ can be given as $\mu_{\mathcal{B}}=0$ and $\sigma_{\mathcal{B}}=N_g \left(K-1\right) \sigma^2_h \sigma^2_{g_k}$, respectively. 
	\end{lemma}\begin{proof}
		 Readers may refer to Appendix B for the derivation steps.
	\end{proof}

Note that $R_k$ and $B$ are the minimum rate at which the quality of service holds for the $k^{\mathrm{th}}$ UE and the bandwidth, respectively. The outage probability of the $k^{\mathrm{th}}$ UE can be found by using the following equation:
\begin{align}
    P_{\mathrm{out}} &= \mathrm{Pr}\left[\gamma_k < 2^{R_k/B}-1 \right], \nonumber \\
     &= \mathrm{Pr}\left[\frac{\left|\mathcal{A}\right|^2}{\left|\mathcal{B}\right|^2 + N_0/P_t} < r_k  \right], \nonumber \\
     &= \mathrm{Pr}\left[\left|\mathcal{A}\right|^2 - r_k \left|\mathcal{B}\right|^2  < \upsilon \right], \nonumber \\   &= \mathrm{Pr}\left[\Upsilon < \upsilon \right] = F_{\Upsilon}(\upsilon) ,
\end{align}
where $r_k = 2^{R_k/B}-1$, $\upsilon = r_k(N_0/P_t)$, and $\Upsilon = \left|\mathcal{A}\right|^2 - r_k \left|\mathcal{B}\right|^2 = \left|\mathcal{A}\right|^2 - \left| \sqrt{r_k} \mathcal{B}\right|^2$ and $F_{\Upsilon}(\upsilon)$ are the difference of independent non-central and central chi-square random variables and the cumulative distribution function of $\Upsilon$, respectively. Moreover, the characteristic function of $\Upsilon$ can be given as \cite{simon2002probability}:
\begin{equation}
    \Psi_{\Upsilon} (\omega) = \exp \left( \frac{j\omega \mu^2_{\mathcal{A}}}{1-2j \omega \sigma^2_{\mathcal{A}} } \right) \frac{1}{ \left(1-2j \omega \sigma^2_{\mathcal{A}} \right)^{0.5}  \left(1+2j \omega \sigma^2_{\mathcal{B}'} \right) },
\end{equation}
where $\sigma^2_{\mathcal{B}} = 0.5 r_k N_g (K-1) \sigma^2_h \sigma^2_{g_k}$ is the variance of $\mathcal{B}' = \sqrt{r_k}\mathcal{B}$ and $\sigma^2_{\mathcal{A}} = N_g  \sigma^2_h \sigma^2_{g_k} \left(1- \frac{\pi^2}{16}\right) $ and $\mu_{\mathcal{A}} = N_g \sigma_h \sigma_{g_k} \frac{\pi}{4}$ are the variance and mean value of $\mathcal{A}$, respectively. Since $\Upsilon$ is the difference of chi-square RVs, obtaining the pdf and cdf of $\Upsilon$ is not straightforward. Therefore, we exploit the Gil-Pelaez's inversion equation \cite{simon2002probability} to obtain the CDF of $\Upsilon$ through characteristic functions as follows:
\begin{equation}
    F_{\Upsilon}(\upsilon) = \frac{1}{2} - \int_{0}^{\infty} \frac{ \Im \left\{e^{-j \omega \upsilon}   \Psi_{\Upsilon}  (\omega) \right\} } {\omega \pi} d \omega.
    \label{eq:cdf1}
\end{equation}

Note that the outage probability can be found by using (\ref{eq:cdf1}).

\subsection{Theoretical Bit Error Rate Analysis}

In this subsection, the theoretical BER of the proposed system is provided. As mentioned in Section III.A, the SINR of the $k^{\mathrm{th}}$ UE ($\gamma_k$) is the ratio of chi-square RVs; thus, one cannot exactly derive the $\gamma_k$ and its closed form does not exist in the literature. Therefore, we exploit the Open Distribution Fitter app (dfittool) of the MATLAB program to create the distribution of $\gamma_k$ as a semi-analytical method. Considering the impact of the interference term $|\mathcal{B}|^2$ on $\gamma_k$ can be neglected, and the chi-square distribution $|\mathcal{A}|^2$ is a variant of the Gamma distribution, $\gamma_k$ is foreseen to follow the Gamma distribution. 

In the context of theoretical derivations, generating shape $(\kappa)$ and scale $(\rho)$ parameters for the gamma distribution corresponding to each configuration is imperative. These parameters are essential components in the analytical framework, facilitating the comprehensive exploration and analysis of the underlying theoretical constructs. As seen from Fig. \ref{fig:distribution}, $\gamma_k$ fits the Gamma distribution for different configurations. With these $\kappa$ and $\rho$ parameters, the PDF of $\gamma_k$ can be given as \cite{lee1979distribution}:
\begin{equation}
    f_{\Gamma} (\gamma_k) = \frac{\gamma^{\kappa-1}_k e^{\left(-\gamma_k/\rho\right)}}{\rho^{\kappa} \Gamma \left(\kappa\right)},
\end{equation}
where $\Gamma (\cdot)$ is the Gamma function. Note that the parameters of the fitted Gamma distribution change for different configurations of UE and $P_t$.

Theoretical symbol error probability (SEP) is computed through the utilization of the moment generation function (MGF) of the Gamma distribution in the following manner:
\begin{equation}
    M_{\gamma_k} (s) = (1-\rho s)^{\kappa}, \quad \text{for } s < \frac{1}{\rho}.
    \label{eq:mgf}
\end{equation}

The average SEP pertaining to $M$-ary PSK modulation, employing the MGF as specified in (\ref{eq:mgf}), is presented as follows:
\begin{equation}
    P_s = \frac{1}{\pi} \int_{0}^{\left(M-1\right)\pi/M} M_{\gamma_k} \left( \frac{-\sin^2\left(\pi/M\right)}{\sin^2 x} \right) dx.
    \label{eq:sep}
\end{equation}

A numerical computation of the SEP can be performed as defined in (\ref{eq:sep}) by employing the parameters of the Gamma distribution. Additionally, bit error probability (BEP) can be approximated as:
\begin{equation}
    P_b \approx P_s / \log_2(M).
    \label{eq:BEP}
\end{equation}

\subsection{RIS Partitioning Optimization}

In the proposed RIS-aided DL NOMA scheme, the RIS is partitioned into equal number of RIS elements. Hence, due to increased path loss, further UEs performs worse than near UEs in terms of many metrics such as error performance and outage probability. This is because when the RIS elements are distributed uniformly, more elements are assigned to near UEs than needed while necessary number of elements not allocated to the further UEs.

In this subsection, we define an optimization problem that considers the outage probability performance of all UEs. In this case, a UE is not assigned an unnecessary number of RIS elements when other users are more in need of those RIS elements hence, all users have a fair power disparity to enable power domain NOMA and make the interference negligible. Here, the optimization problem (P1) is defined in such a way that more RIS elements are allocated to further UEs in order to obtain outage probability performance in a more fair manner as follows:

\begin{align}
    \left(\mathrm{P1}\right) = &\underset{\mathbb{N}}{\mathrm{min}} \sum_{k=1}^{K} \sum_{k'=k+1}^{K} \left|P^{(k)}_{\mathrm{out}} - P^{(k')}_{\mathrm{out}} \right|, 
    \nonumber \\
        &\text{s.t.}\hspace*{0.2cm} \sum_{k=1}^{K} N^{(k)} = N.
\end{align}
where $\mathbb{N}=\{N^{(1)}, N^{(2)}, \cdots, N^{(K)}\}$ and $N^{(k)}$ is the number of RIS elements allocated to the $k^{\mathrm{th}}$ UE.

We utilize a simple search algorithm to solve $(\mathrm{P1})$ for $K=2$. Here, the value of $N^{(1)}$ is increased from $1$ to $N$ with an an increment of $1$, while $N^{(2)}$ is equal $N-N^{(1)}$. When $K>2$, this algorithm becomes significantly complex; therefore, a low-complexity algorithm to solve $(\mathrm{P1})$ for $K>2$ is left as a future study item.

\section{Simulation Results And Comparisons}   

In this section, we provide the computer simulation results of the proposed scheme and compare them to the OMA benchmark. Furthermore, each figure for the proposed scheme and benchmark is confirmed with theoretical simulations. Performances of the outage probability, QoS, sum-rate, and BER are shown and discussed for varying parameters.

\subsection{Benchmark Schemes}
In this study, time division multiple access (TDMA) is considered as a benchmark scheme for performance comparison. To find the outage probability of the TDMA, its SNR expression for the $k^{\mathrm{th}}$ UE is needed and can be given as follows, respectively:
\begin{align}
    \gamma^{\mathrm{TDMA}}_k = \frac{P_t \left|\Bar{\mathcal{A}}\right|^2}{N_0},
    \label{fig:SNR_TDMA}
\end{align}
where $\Bar{\mathcal{A}}=\sum_{n\in \mathbb{I}}  \alpha_n \beta_{n,k}$ and $\mathbb{I} = \{1,\cdots,N\}$ is a set of indices of all RIS elements. Since all elements of the RIS serve only one UE at a given time slot, there is no interference term in (\ref{fig:SNR_TDMA}). It should be noted that for a fair comparison, since the UEs do not share the same time resource and communicate simultaneously, the time allocated for each UE is divided into $K$; hence, the rate of data for each UE is multiplied by $K$. Finally, the outage probability of the $k^{\mathrm{th}}$ UE for TDMA is given as:
\begin{equation}
     P^{\mathrm{TDMA}}_{\mathrm{out},k} = \mathrm{Pr}\left[\gamma^{\mathrm{TDMA}}_k < 2^{\left(K R_k/B\right)}-1\right].
\end{equation}

Additionally, the sum rate for TDMA can be expressed as $(\sum_{k} \gamma^{\mathrm{TDMA}}_k)/K$.

\subsection{Simulation Setup Parameters}

In this subsection, the simulation setup parameters are provided. The $k^{\mathrm{th}}$ UE, an RIS, and a BS are located in 2D coordinate system with coordinate vectors $\mathbf{p}^{\mathrm{UE}_k} = [x^{\mathrm{UE}_k},y^{\mathrm{UE}_k}]^{\mathrm{T}}$, $\mathbf{p}^{\mathrm{RIS}} = [x^{\mathrm{RIS}},y^{\mathrm{RIS}} ]^{\mathrm{T}}$, and $\mathbf{p}^{\mathrm{BS}} = [ x^{\mathrm{BS}},y^{\mathrm{BS}}]^{\mathrm{T}}$, respectively, $k=1,\cdots,K$. Hence, the distance between the $k^{\mathrm{th}}$ UE and the RIS and the RIS and BS can be calculated as $d^{{\mathrm{UE}_k} \mathrm{-RIS}} = || \mathbf{p}^{\mathrm{UE}_k} - \mathbf{p}^{\mathrm{RIS}} ||$ and $d^{{\mathrm{RIS}} \mathrm{- BS}} = || \mathbf{p}^{\mathrm{RIS}} - \mathbf{p}^{\mathrm{BS}} ||$, respectively. In order to calculate the path loss with the given distances, the third generation partnership project 
 (3GPP) UMi path loss model for non-line-of-sight (NLOS) transmission defined within the $2$-$6$ GHz frequency band and Tx-Rx range varying from $10$-$2000$ m, can be given as:
\begin{equation}
    L\left(d^i\right) \: \left[\mathrm{dB}\right] = 36.7 \log_{10}\left(d^i\right) + 22.7 + 26 \log_{10} \left(f_c\right),
\end{equation}
where $i\in\{\mathrm{UE}_k \mathrm{-RIS}, \mathrm{RIS-BS}\}$ and $f_c$ is the operating frequency. The power of the channel between the $k^{\mathrm{th}}$ UE and RIS and the RIS and BS can be calculated as: $\sigma^2_{g_k} = 1/10^{L(d^{\mathrm{UE}_k-\mathrm{RIS}})/10}$ and $\sigma^2_{h} = 1/10^{L(d^{\mathrm{RIS-BS}})/10}$, respectively. The parameters considered in computer simulations are given in Table \ref{table:sim_param}. Note that the transmit power of the BS ($P_t$) is in the range of  $-30$ and $40$ dBm, and and operating frequency ($f_c$) is $2.4$ GHz as specified \cite{3gpp2019nr}. In addition, RIS element sizes are in the range of $64$-$1024$ as can be confirmed in \cite{araghi2022reconfigurable}.

\renewcommand{\arraystretch}{1.5}
\begin{table}[t]
\caption{Computer Simulation Parameters}
\centering
\label{table:sim_param}
\begin{tabular}{|l|l|l|}
\hline
 Carrier frequency (GHz) & $f_c$                    & $2.4$                 \\ \hline
Location of UE1 & $\mathbf{p}^{\mathrm{UE}_1}$ & $[15,12]^{\mathrm{T}}$ \\ \hline
Location of UE2 & $\mathbf{p}^{\mathrm{UE}_2}$ & $[10,8]^{\mathrm{T}}$  \\ \hline
Location of UE3 & $\mathbf{p}^{\mathrm{UE}_3}$ & $[5,2]^{\mathrm{T}}$   \\ \hline
Location of UE4 & $\mathbf{p}^{\mathrm{UE}_4}$ & $[2,1]^{\mathrm{T}}$   \\ \hline
Location of BS & $\mathbf{p}^{\mathrm{BS}}$   & $[55,10]^{\mathrm{T}}$ \\ \hline
Location of RIS & $\mathbf{p}^{\mathrm{RIS}}$  & $[35,5]^{\mathrm{T}}$  \\ \hline
Noise variance (dBm) & $N_0$                    & $-130$                \\ \hline
\end{tabular}
\end{table}

\begin{figure}[t]
    \centering
    \includegraphics[width=0.85\columnwidth]{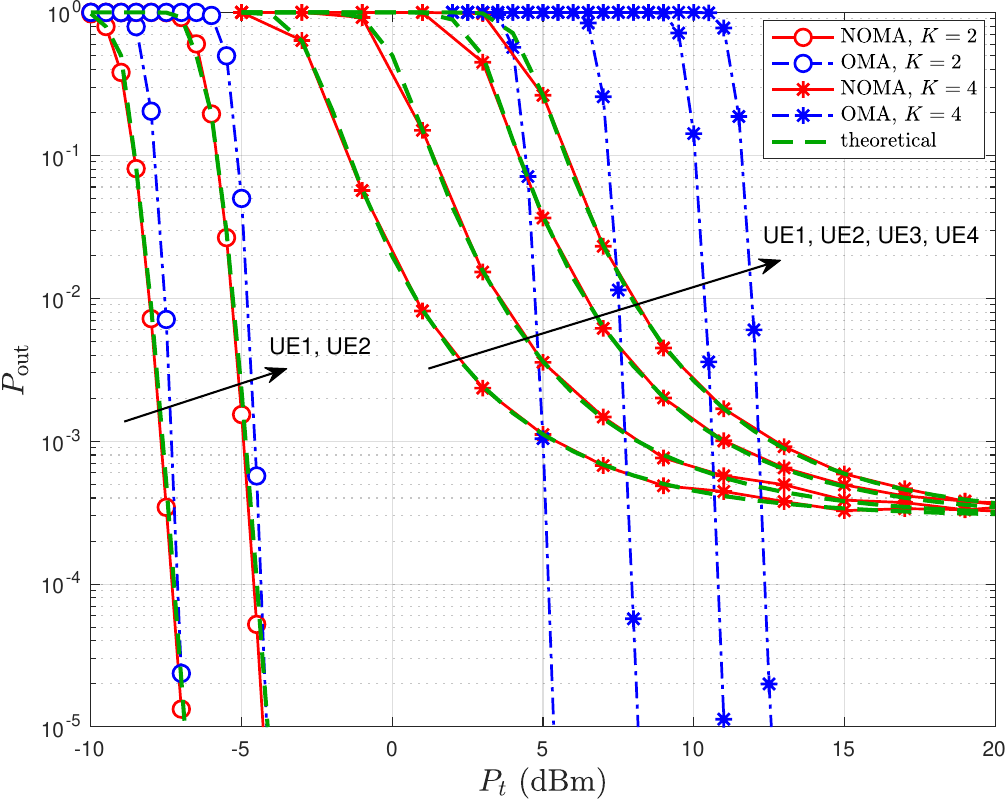}
    \vspace{-0.2cm}
    \caption{Comparison of the outage probability performance of the proposed DL-NOMA scheme and TDMA for $R_k=2$ bps/Hz, $N=512$ and $K=2$, $4$.}
    \label{fig:Pout1}
\end{figure}

\subsection{Computer Simulations and Discussions}

In this subsection, comprehensive simulation results are presented to investigate the performance of the proposed DL-NOMA scheme. We investigate the performance of outage probability (${P}_{\mathrm{out}}$), sum rate, quality-of-service (QoS), and BER for varying parameters, cases and UEs. Additionally, the performance of the proposed DL-NOMA is compared to the TDMA-OMA benchmark scheme. As it can be seen in all simulation results, the theoretical results align and match exactly in performance and behaviour. For all simulations, the computer simulation set-up parameters in Table I are considered.

\begin{figure}[t]
    \centering
    \includegraphics[width=0.85\columnwidth]{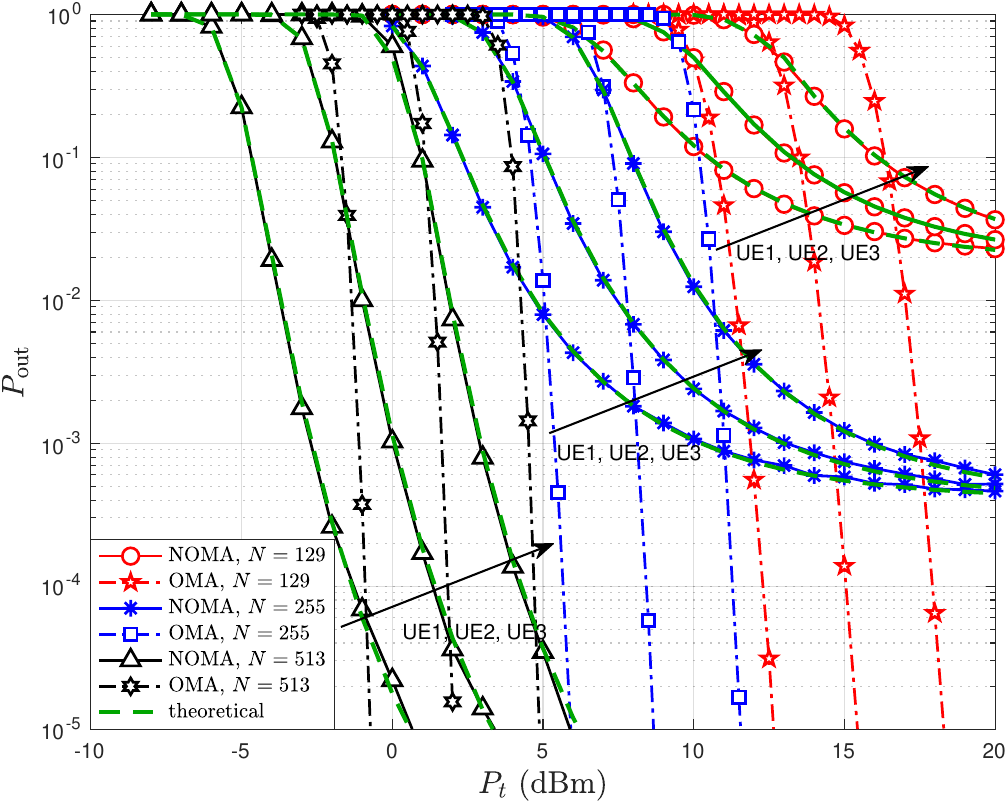}
    \vspace{-0.2cm}
    \caption{Comparison of the outage probability performance of the proposed DL-NOMA scheme and TDMA for $R_k=2$ bps/Hz, varying $N$ and $K=3$.}
    \label{fig:Pout2}
\end{figure}

\begin{figure}[t]
    \centering
    \includegraphics[width=0.85\columnwidth]{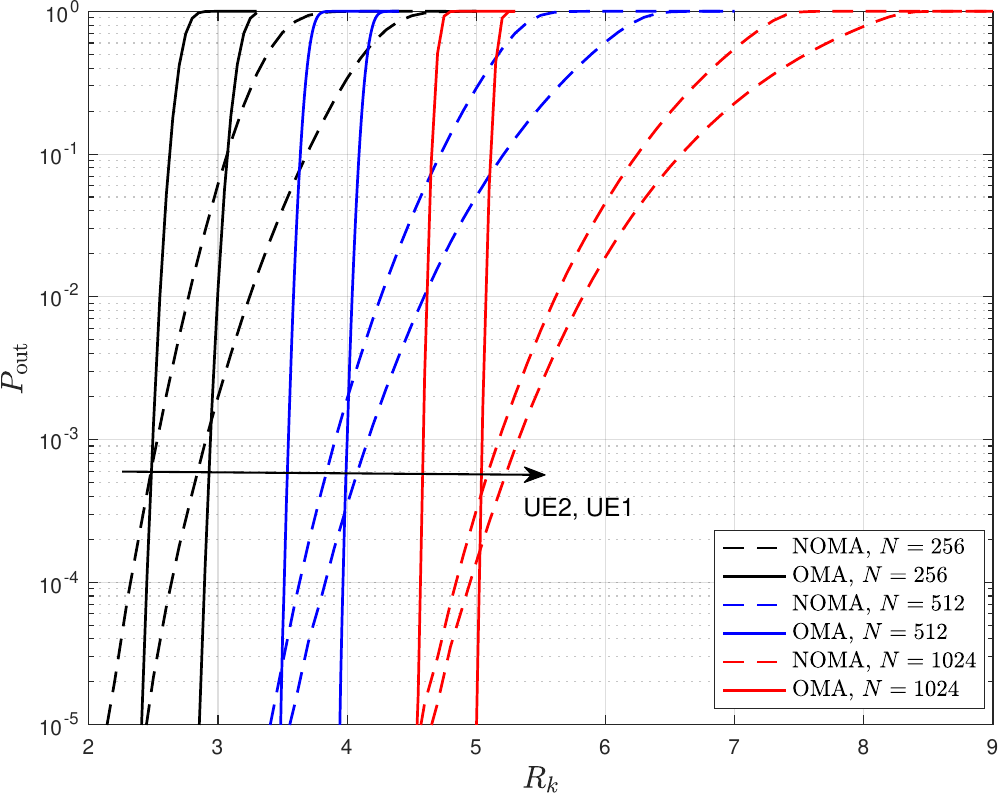}
    \vspace{-0.2cm}
    \caption{Outage probability vs QoS performance comparison of the proposed DL-NOMA and OMA schemes for $K=2$, varying $N$, $P_t = 5$ dBm.}
    \label{fig:Qos}
\end{figure}

\begin{figure}[t]
    \centering
    \includegraphics[width=0.85\columnwidth]{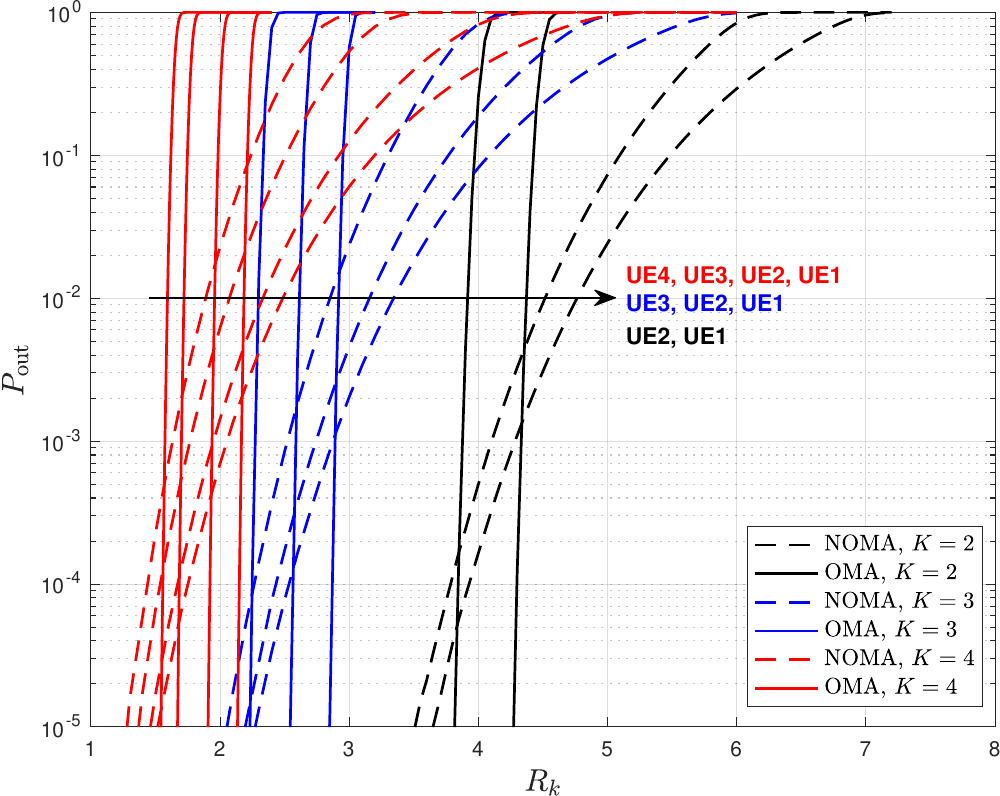}
    \vspace{-0.2cm}
    \caption{Outage probability vs QoS performance comparison of the proposed DL-NOMA and OMA schemes for varying $K$, $N=512$, $P_t = 7$ dBm}
    \label{fig:Qos2}
\end{figure}

In Figs. \ref{fig:Pout1} and \ref{fig:Pout2}, the outage probability performance of the proposed DL-NOMA scheme is presented and compared to TDMA OMA benchmark. In Fig. \ref{fig:Pout1}, the parameters are selected as $N=512$ and $K=2, 4$. As seen from Fig. \ref{fig:Pout1}, the performance of the proposed system is better in the case of $K=2$. Since the interference effect is more severe for the scenario with more UEs, the outage probability performance is better when $K=2$ compared to the case where $K=4$. Since there is an increase interference in the case of $K=4$, an error floor is observed at around $P_{out}=10^{-4}$. When $K=4$, the proposed DL-NOMA scheme outperforms our benchmark until it reaches a certain point of saturation in terms of outage probability performance for all UEs; however, it becomes worse than the benchmark as $P_t$ increases due to the interference effect of the UEs which can also be observed by the error floor. It can also be seen that the theoretical curves derived from (\ref{eq:cdf1}) verify our simulation results.

In Fig. \ref{fig:Pout2}, the outage probability performance of the proposed DL-NOMA scheme is demonstrated and compared with TDMA for $K=3$ and varying RIS sizes $N$. As seen from Fig. \ref{fig:Pout2}, as $N$ increases, the outage probability performance of the proposed DL-NOMA scheme significantly improves due to the gains from the RIS elements. For small values of $N$, the interference of the UEs has a higher impact which can be seen by the error floors and the performance of the proposed scheme is less impactful. Therefore, the superiority of the proposed DL-NOMA scheme over the benchmark in terms of outage probability is more obvious when $N$ is determined as a high value because the interference of the UEs becomes more negligible and the RIS can enable NOMA more effectively. Furthermore, it can be seen that the theoretical curves obtained by (\ref{eq:cdf1}) coincide with the simulation results for different $N$ values.

\begin{figure}[t]
    \centering
    \includegraphics[width=0.85\columnwidth]{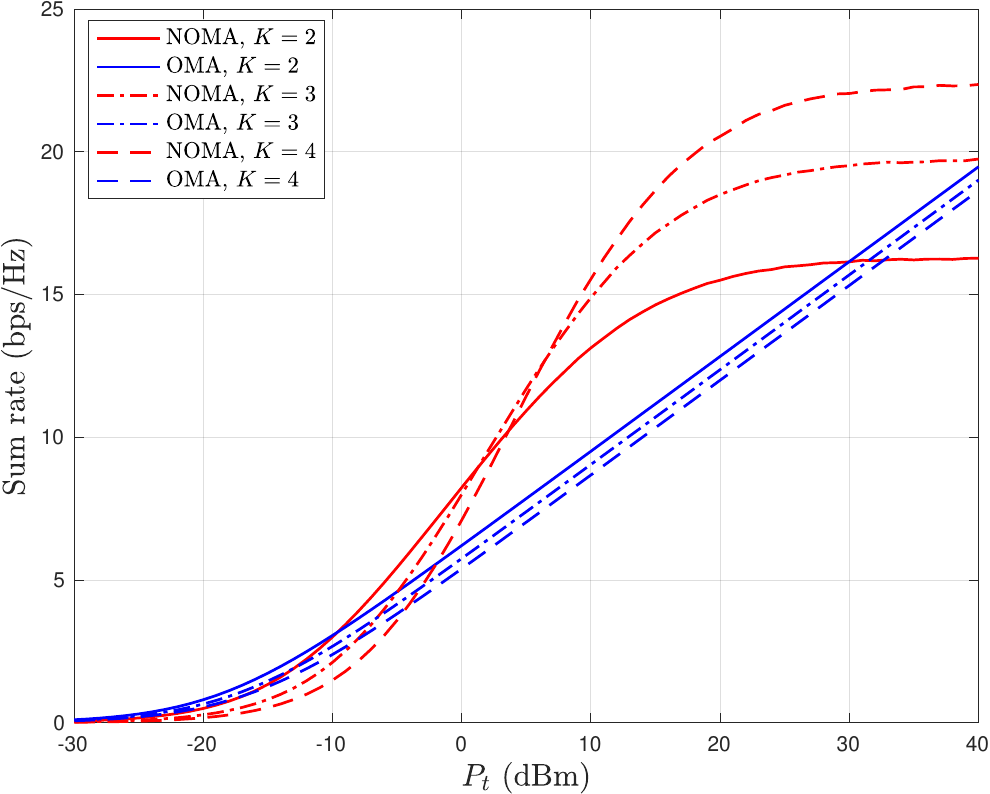}
    \vspace{-0.2cm}
    \caption{Sum rate comparison of the proposed DL-NOMA and OMA schemes when $N=512$ for varying $K$.}
    \label{fig:sumrateONE}
\end{figure}

In Fig. \ref{fig:Qos}, the performance superiority of the proposed scheme over the benchmark can be observed as the QoS requirement becomes more strict. As the QoS requirement for a UE's communication performance gets higher, the outage probability increases since it is harder to sustain that quality. This can be seen in Fig. \ref{fig:Qos} for both UEs and varying RIS size. 
It should be noted that, while the benchmark outperforms the proposed NOMA scheme for low QoS value, after a certain QoS value,  the proposed scheme shows its superiority over OMA. Additionally, at higher QoS values, NOMA enables communication to the users whereas communication can not be longer sustained in the OMA case.  It can be seen that as the RIS size increases, the outage probability performance improves for both UEs. It should be noted that in NOMA the UEs QoS performance is almost similar whereas there is a larger difference in performance in the OMA benchmark. Furthermore, it can be analyzed that as the RIS size ($N$) increases, the system can sustain higher QoS requirement. Since UE1 is closer to the BS compared to UE2, its performance is better.

In Fig. \ref{fig:Qos2}, the outage probability performance of the proposed DL NOMA scheme is demonstrated with respect to varying QoS ($R_k$) values for $P_t = 7$ dBm and $B=1$ where  $N=512$ and $513$ for $K=2$, $4$, and $3$, respectively. The reason why $N$ is selected as $513$ for $K=3$ is to allocate the same integer number RIS elements to each UE. Additionally, the outage probability performance of DL NOMA scheme is compared to OMA. As seen from Fig. \ref{fig:Qos2}, as the number of UEs ($K$) decreases, higher QoS values can be provided because more RIS elements can be allocated to a specific UE for a given RIS size ($N$) in a system with fewer UEs. When the proposed DL NOMA scheme is compared to OMA, it can be seen that after a certain QoS value, the OMA scheme can no longer enable communication while the proposed scheme can for all values of $K$.

\begin{figure}[t]
    \centering
    \includegraphics[width=0.85\columnwidth]{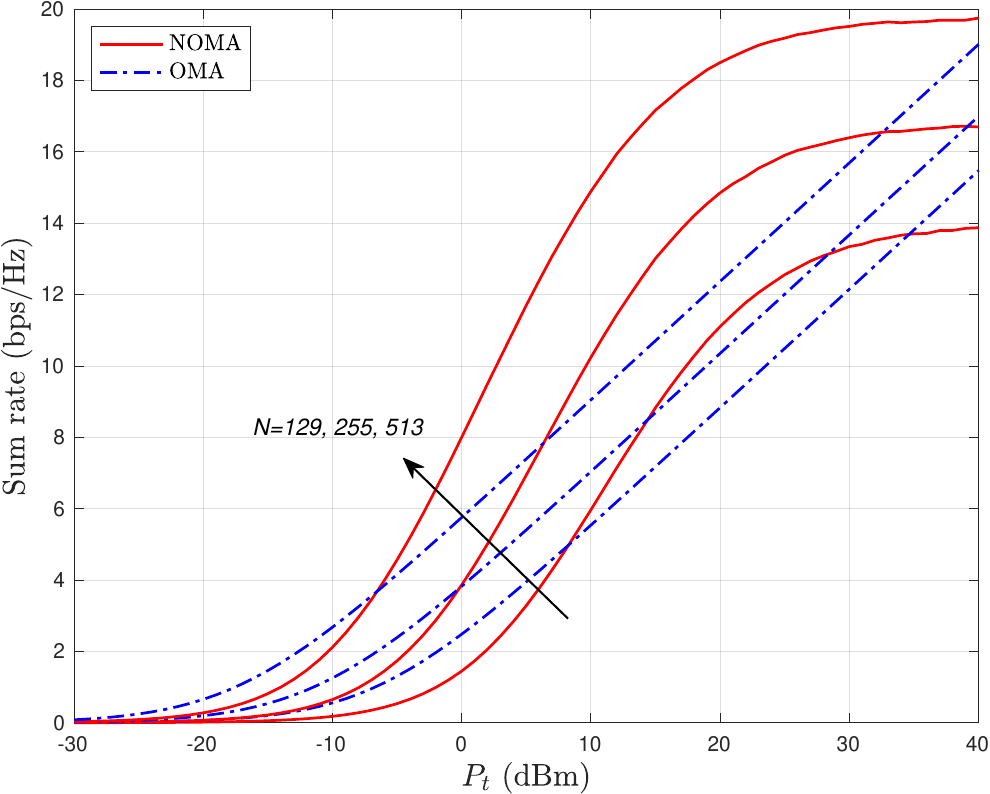}
    \vspace{-0.2cm}
    \caption{Sum rate comparison of the proposed DL-NOMA and OMA schemes for $K=3$ and varying $N$ values.}
    \label{fig:sumrateTWO}
\end{figure}

\begin{figure}[t]
    \centering
    \includegraphics[width=0.85\columnwidth]{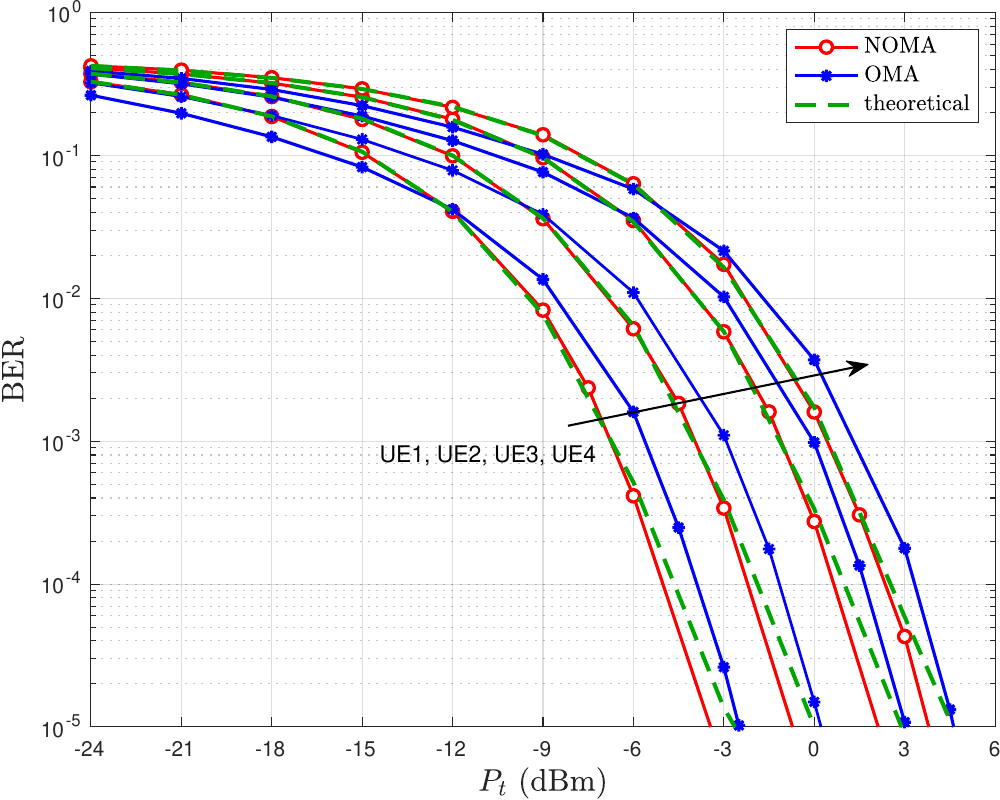}
    \vspace{-0.2cm}
    \caption{BER performance comparison of the proposed DL-NOMA and OMA schemes for $N=1024$, $K=4$, and $1$ bps/Hz.}
    \label{fig:BER_ONE}
\end{figure}

\begin{figure}[t]
    \centering
    \includegraphics[width=0.85\columnwidth]{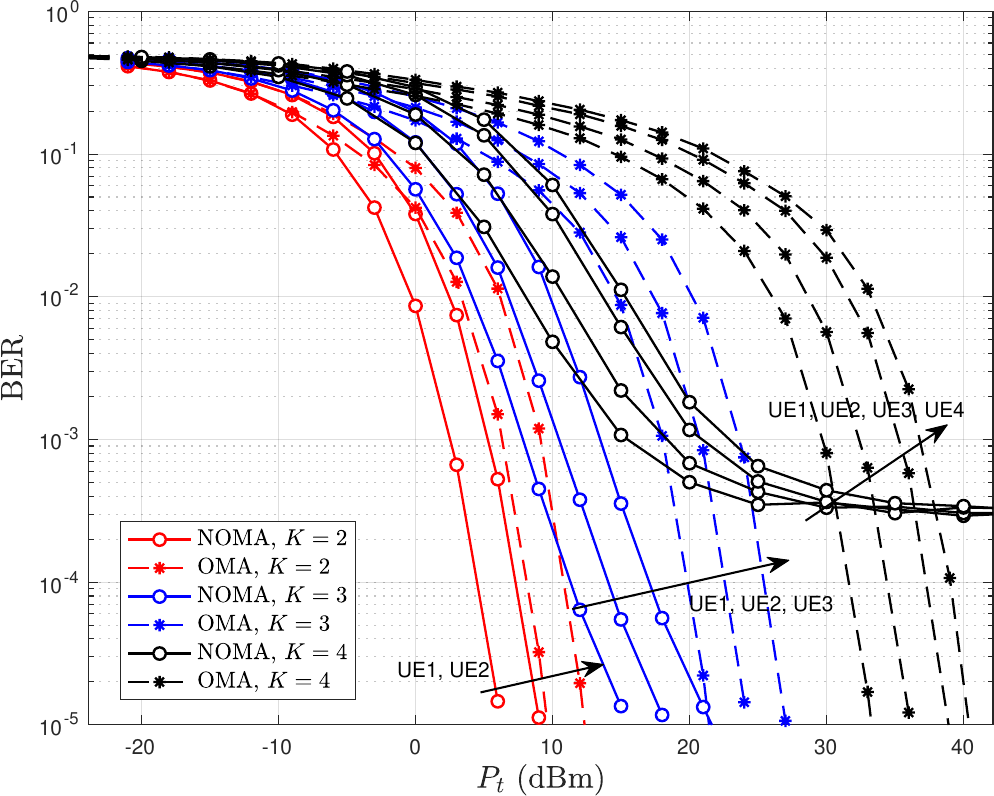}
    \vspace{-0.2cm}
    \caption{BER performance comparison of the proposed DL-NOMA and OMA schemes for $N=256$, $2$ bps/Hz, and varying number of UEs ($K$).}
    \label{fig:BER_TWO}
\end{figure}

In Figs. \ref{fig:sumrateONE} and \ref{fig:sumrateTWO}, the sum rate superiority of the proposed scheme over the benchmark can be analyzed as the transmit power of the system increases for varying UEs. It can be visualized in Fig. \ref{fig:sumrateONE} that as $P_t$ increases the TDMA benchmark also increases linearly and an increase in the number of UEs does not introduce immense gains in the system. On the other hand, the proposed scheme's sum rate significantly increases and surpasses that of the benchmark. Furthermore, an increase in the number of UEs in the NOMA system has a notable positive impact on the systems sum rate. In Fig. \ref{fig:sumrateTWO}, the sum rate performance is considered as the number of RIS elements $N$ increases. It can be seen that as before, an increase in $P_t$ has a great impact on improving the proposed system performance, while for the benchmark it only provides a linear gain. Additionally, it can also be observed that for a small $N$ the proposed system still outperforms the benchmark however, the proposed system becomes much more effective and the performance gains are significantly enhanced for a larger $N$.

\begin{table*}[th]
\caption{Fitted Gamma Distribution Parameters for Different UEs and Varying $P_t$. \vspace{-0.4cm}} % title name of the table 
\centering % centering table
\scalebox{0.8}{\begin{tabular}{l c r rrrrrrrrrr} % creating 10 columns
\\
\hline 
 UE & $P_t$ (dB): &  -24 & -21 & -18 & -15 & -12 & -9 & -6 & -3 & 0 
\\ [0.5ex]
\hline \hline % inserts single-line
% Entering 1st row
& $\kappa$    & 102.558  & 102.762 & 102.741  & 101.213 &  94.963 & 79.924 & 52.903 & 27.465 & 12.893 \\[.5ex]
 UE 1 & $\rho$    & 0.0009746 &   0.001937 & 0.0038528 &  0.0077486  & 0.0162433  & 0.03749 & 0.107683 & 0.38155 & 1.42646 \\[2ex]
 \hline
% Entering 2nd row
& $\kappa$     & 103.181 &  102.873  & 102.771  &   102.713  & 100.714  & 94.803 & 78.058 & 50.728 & 25.909   \\[.5ex]
  UE 2 & $\rho$   &    0.00051472  &  0.00102934  &  0.00205164 &  0.00407926  & 0.00823733 &  0.0172064 & 0.0405408 & 0.118369 & 0.424852 \\[2ex]
 \hline
% Entering 3rd row
   & $\kappa$     & 103.448 &  103.288  & 103.323 & 102.947  & 102.796 & 100.473 & 95.084 & 77.188 & 49.051 \\[.5ex]
 UE 3  & $\rho$    &  0.00026515 & 0.000529542 &  0.00105591 & 0.00210947 & 0.00419777 & 0.00850284 & 0.0176538 & 0.0421751 & 0.125799 \\[2ex]
 \hline
 %enter 4th row
   & $\kappa$     & 103.151 &  103.293  & 102.924  &  103.789  & 103.372  & 101.781 & 98.843 & 87.664 & 64.441 \\[.5ex]
UE 4 & $\rho$  &    0.00018594 &   0.000370327 &  0.000740912 &  0.00146360 & 0.00292481 &  0.00589186 & 0.0119778 & 0.0263828 & 0.06889412 \\[1ex]
% [1ex] adds vertical space
\hline % inserts single-line
\end{tabular}}
\label{tab:PPer}
\end{table*}

\begin{figure}[t]
    \centering
    \includegraphics[width=0.85\columnwidth]{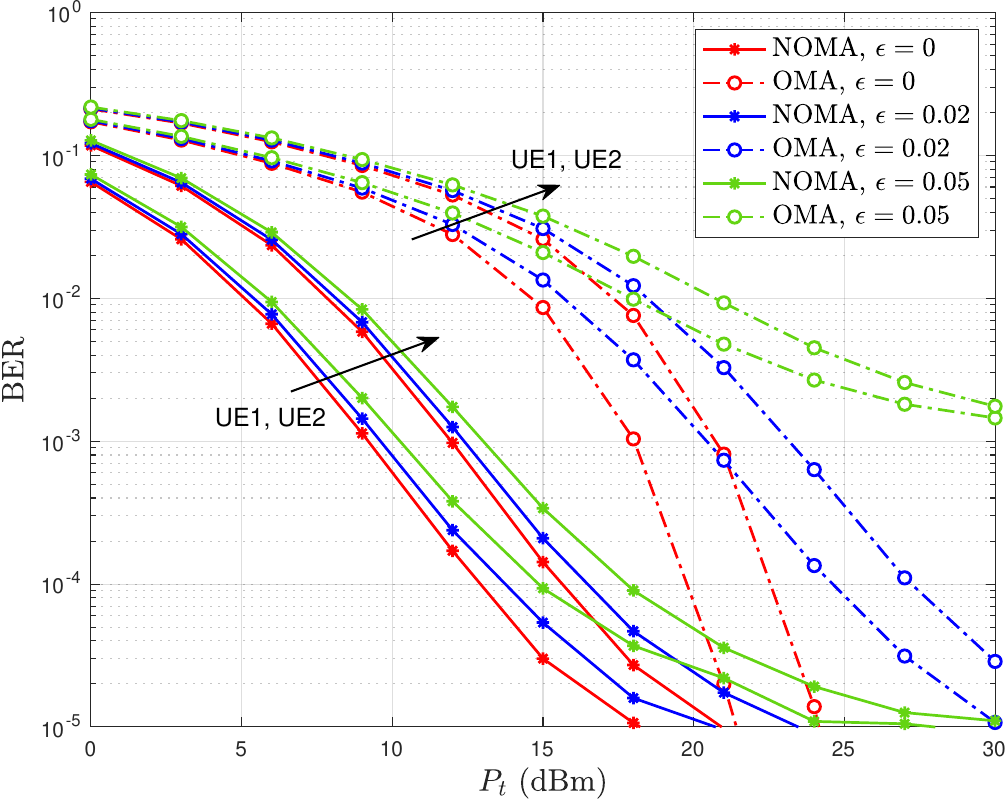}
    \vspace{-0.2cm}
    \caption{BER performance comparison of the proposed NOMA and OMA schemes under imperfect CSI knowledge for $3$ bps/Hz, $N=256$, $M=4$, and $K=2$.}
    \label{fig:BER_Three}
\end{figure}

Figs \ref{fig:BER_ONE}, \ref{fig:BER_TWO} and \ref{fig:BER_Three} provide BER comparisons of the proposed system and TDMA benchmark for a varying $P_t$. In Fig. \ref{fig:BER_ONE}, BER performance of the proposed DL NOMA scheme is compared to OMA for $K=4$, $N=1024$, and $1$ bps/Hz. Note that ML detection rule in (\ref{eq3}) is used to decode the DL NOMA signal. In order to provide $1$ bps/Hz, $M$ is selected as $2$ and $16$ for NOMA and OMA schemes, respectively. It can be observed that the proposed NOMA outperforms the OMA benchmark for all UEs. Furthermore, the nearest UE to BS (UE1) provides the best error performance and the furthest UE to BS (UE4) has the worst as expected. Moreover, it can be seen that for low $P_t$ values, the theoretical results obtained by using (\ref{eq:BEP}) and the parameters in Table \ref{tab:PPer} match the simulated BER results. However, after a certain $P_t$ value, a gap between theoretical and simulation curves emerges and, this gap increases with respect to $P_t$ since the SINR of the $k^{\mathrm{th}}$ UE ($\gamma_k$) no longer follows the Gamma distribution exactly due to the non-negligible interference effect.

On the other hand, in Fig. \ref{fig:BER_TWO}, the BER performance vs transmit power ($P_t$) of the proposed scheme and benchmark is shown for a varying number of UEs ($K$), $N=256$, and $2$ bps/Hz. Here, $M$ is chosen  as $4$ and $4^K$ for NOMA and OMA, respectively, to provide $2$ bps/Hz spectral efficiency. Note that $N$ is selected as $255$ for $K=3$ since $256/3$ is not an integer. Similarly in the BER performance, the case of $K=2$ provides the best performance and an increase in $K$ introduces a degradation in the system performance for the proposed NOMA since interference between UEs becomes more dramatic. For $K=2$ and $3$, it can be seen that NOMA scheme outperforms OMA for all UEs. Additionally, for $K=4$, NOMA outperforms OMA for $P_t$ values less than $30$ dBm due to lower modulation order. However, for $P_t$ values greater than approximately $30$ dBm, NOMA shows worse error performance than OMA since interference causes an error floor in BER curve.

In Fig. \ref{fig:BER_Three}, the effect of imperfect channel knowledge at RIS over error performance is investigated for DL NOMA scheme for $3$ bps/Hz, $N=256$ and $K=2$. Note that $M$ is selected as $8$ and $8^K$ for NOMA and OMA, respectively. In order to model the imperfect channel estimation, $\theta_n$ is obtained by using $\hat{\phi}_n$ and $\hat{\psi}_{n,k}$ instead of $\phi_n$ and $\psi_{n,k}$, where $\hat{\phi}_n = \angle \hat{h}_n$, $\hat{\psi}_{n,k} = \angle \hat{g}_{n,k}$, $\hat{h}_n = h_n - \epsilon_h$, $\hat{g}_{n,k} = g_{n,k} - \epsilon_g$, $\epsilon_h \sim \mathcal{CN}\left(0, \zeta \sigma^2_h\right)$, and $\epsilon_g \sim \mathcal{CN}\left(0, \zeta \sigma^2_{g_k}\right)$. The BER simulations in Fig. \ref{fig:BER_Three} have been realized under $\zeta=0.05$, $\zeta=0.02$ and perfect CSI ($\zeta=0$). It can seen from Fig. \ref{fig:BER_Three} that channel estimation errors impact the benchmark scheme more dramatically compared to the proposed scheme. Additionally, NOMA scheme outperforms OMA for all cases.

\begin{figure}[t]
    \centering
    \includegraphics[width=0.85\columnwidth]{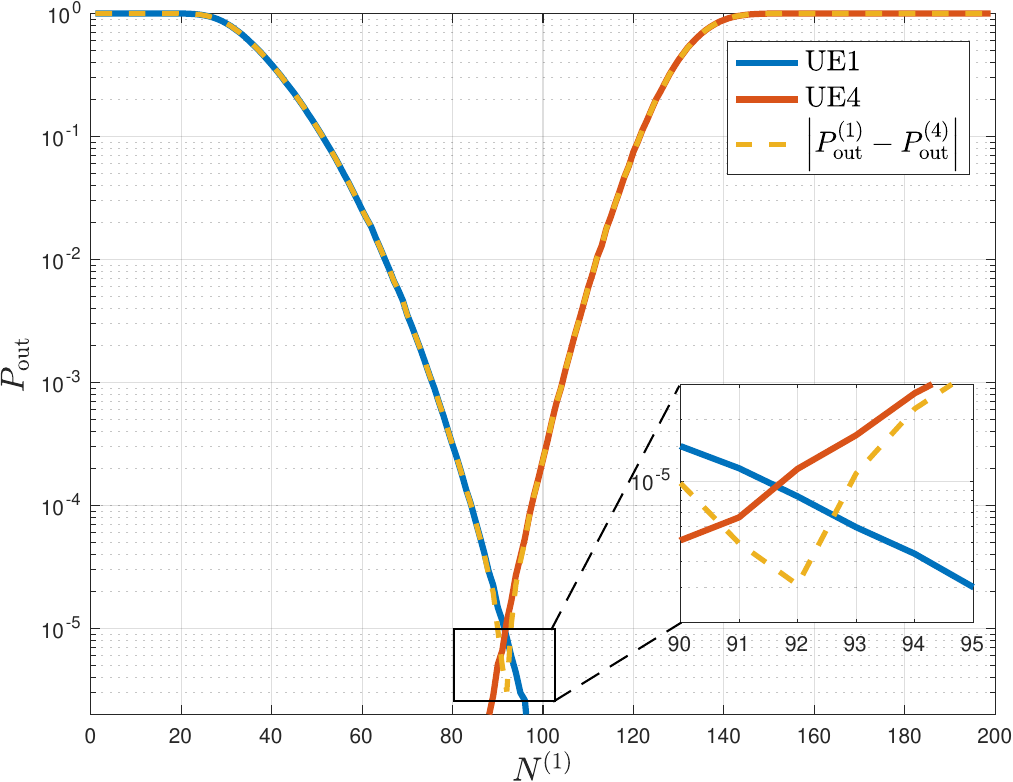}
    \vspace{-0.2cm}
    \caption{Determining the value of $N^{(1)}$ to solve $(\mathrm{P1})$ for $K=2$, $P_t = 10$ dBm, and $N=200$.}
    \label{fig:Subfigure_1}
\end{figure}

\begin{figure}[t]
    \centering
    \includegraphics[width=0.85\columnwidth]{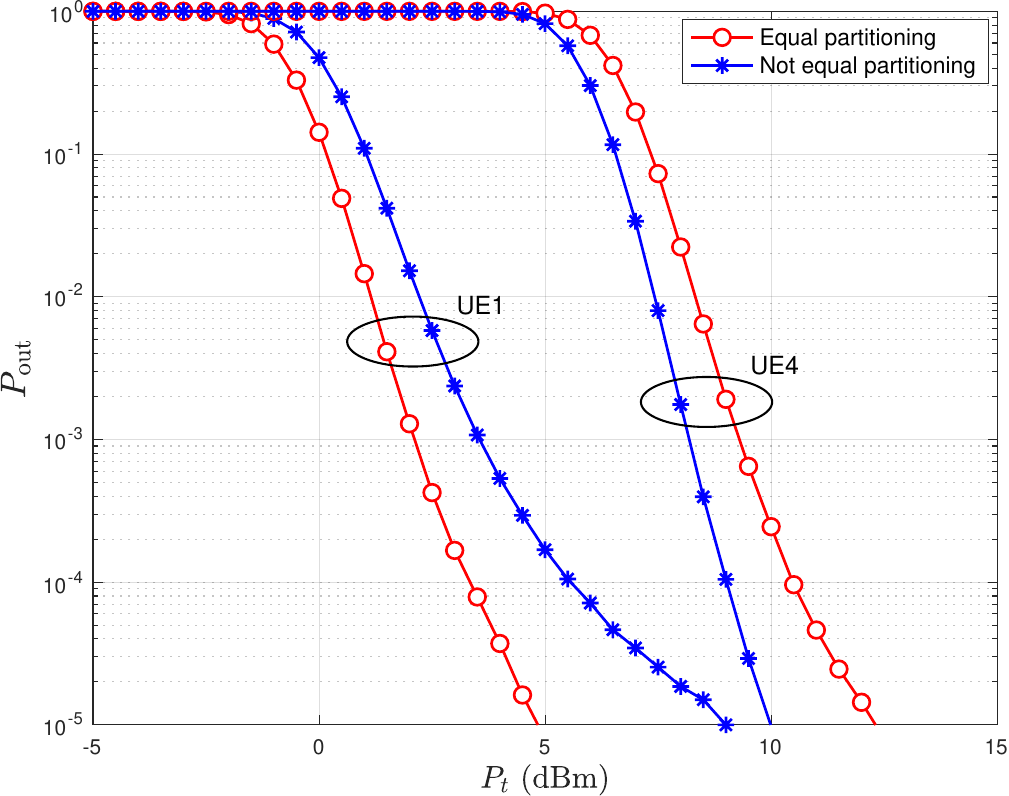}
    \vspace{-0.2cm}
    \caption{Outage probability performance comparison of equal partitioning and not equal partitioning with $N^{(1)}=92$ for $K=2$, $N=200$, and $R_k=2$ bps/Hz.}
    \label{fig:Subfigure_2}
\end{figure}

In Fig. \ref{fig:Subfigure_1}, the outage probability performance is investigated for the two most distant UEs (UE$1$ and UE$4$) when $N^{(1)}$ is increased from $1$ to $N-1$ with an increment of $1$, where $N=200$ and $P_t=10$ dBm. As seen from Fig. \ref{fig:Subfigure_1}, $N^{(1)}$ is obtained as $92$ which minimizes $\left| P^{(1)}_{\mathrm{out}} - P^{(4)}_{\mathrm{out}}  \right|$ , where $N^{(4)}=N-N^{(1)}=108$. From the fairness perspective, since UE$1$ is closer to the BS and has better channel conditions and UE$4$ is further to the BS with worse channel conditions, it makes sense to allocate more RIS elements to UE$4$. This is also confirmed with the proposed sub-optimal algorithm and we can see this in Fig. \ref{fig:Subfigure_2}, where we present the the outage probability performance comparison for the scenarios where RIS has uniform and sub-optimal partitioning for $K=2$, $N=200$, and $R_k=2$ bps/Hz. $N^{(1)}$ is selected as $92$ as in \ref{fig:Subfigure_2}. As we can observe, the performance of UE$4$ significantly improves which there is a trade-off from the performance of UE$1$. Most importantly, we can observe that the outage probability curves reach closer together for both UEs hence, providing a more fair system and RIS element partitioning.

%\begin{figure}[t]
%		\centering
%		\subfloat[][]{\includegraphics[width=0.65\columnwidth]{SubFigure1.pdf}}
%		\hfilx
%		\subfloat[][]{\includegraphics[width=0.65\columnwidth]{SubFigure2.pdf}}
 %           \hfil
%		\caption{ (a)  (b) .}
%		\label{Subfigure}
%	\end{figure}

\section{Conclusion}	
In this paper, we have proposed a novel over-the-air DL NOMA scheme with the aid of an passive RIS. In addition, a thorough theoretical analysis considering outage probability and BER is presented along with computer simulation results and each are discussed along with a benchmark comparison. Furthermore, additional analysis such as sum-rate and consideration of imperfect CSI conditions are included. While the NOMA process is conventionally conducted at the BS or UEs, the proposed system enables NOMA through the RIS and virtually enabling the NOMA process over-the-air. Furthermore, SIC is a major concern in most NOMA systems where errors in SIC reduce the system performance and the entire SIC process uses energy and increases complexity of the system. However, in the proposed system, SIC is not required hence reducing complexity at the UEs and BS and reduces the possibility of SIC errors in degrading the system performance. Future works may consider the incorporation of active RIS structure, RIS element optimization, use of deep learning models, further theoretical analysis regarding different channel models such as Rician fading and much more. 

	\balance

%\newpage

	\appendices
	\numberwithin{equation}{section}
	\section{Proof of Lemma \ref{lemma:OPA}}
	\label{AppOP}
Let us define 
\begin{align}
    \label{eq:Lemma1_eq1}
    \mathcal{A} =\sum_{n\in \mathbb{I}_k}\left|h_n\right|\left|g_{n,k}\right| =  \sum_{n\in \mathbb{I}_k}  \alpha_n \beta_{n,k},
\end{align}
where $\mathbb{I}_k = \left\{(k-1)N_g + 1, (k-1)N_g + 2, \cdots, k N_g \right\}$ is a set of RIS elements indices allocated for the $k^{\mathrm{th}}$ UE. Since the phase value of $n^{\mathrm{th}}$ RIS element is determined to provide coherent alignment with the corresponding channel coefficients $h_n$ and $g_{n,k}$ for $n \in \mathbb{I}_k$, the phase values of these coefficients are removed, causing the multiplication of amplitudes only as in (\ref{eq:Lemma1_eq1}). Note that $\alpha_n$ and $\beta_{n,k}$ are i.i.d. Rayleigh distributed random variables, and the mean of the multiplication of them can be given as:
 \begin{align}
    \label{eq:Lemma1_eq2}
     \mathrm{E}\left[\alpha_n \beta_{n,k} \right]=\mathrm{E}\left[\alpha_n\right]\mathrm{E}\left[\beta_{n,k}\right] = \frac{\sigma_{h}\sigma_{g_{k}}\pi}{4},
 \end{align}
 where $\mathrm{E}\left[\alpha_n\right] = \sigma_h \sqrt{\pi/4}$ and $\mathrm{E}\left[\beta_{n,k}\right] = \sigma_{g_k} \sqrt{\pi/4}$.

Additionally, the variance can be given as:
 \begin{align}
 \label{eq:Lemma1_eq3}
     \mathrm{Var}\left[\alpha_n \beta_{n,k}\right]=\mathrm{E}\left[\left[\alpha_n \beta_{n,k}\right]^2 \right]-\left(\mathrm{E}\left[\alpha_n \beta_{n,k}\right] \right)^2.
 \end{align}
By taking square of (\ref{eq:Lemma1_eq2}), $\left(\mathrm{E}\left[\alpha_n \beta_{n,k}\right] \right)^2$ can be calculated  as
\begin{equation}
\label{eq:Lemma1_eq4}
    \left(\mathrm{E}\left[\alpha_n \beta_{n,k}\right] \right)^2 = \frac{\sigma^2_{h}\sigma^2_{g_{k}}\pi^2}{16}.
\end{equation}

Since $\alpha_n$ and $\beta_{n,k}$ are i.i.d. random variables, we have the following derivations:
\begin{align}
\label{eq:Lemma1_eq5}
    \mathrm{E} \left[ \left[\alpha_n \beta_{n,k}\right]^2 \right] = \mathrm{E} \left[ \left[\alpha^2_n \beta^2_{n,k}\right] \right] = \mathrm{E} \left[ \alpha^2_n \right] \mathrm{E} \left[ \beta^2_{n,k} \right].
\end{align}

In order to find $\mathrm{E} \left[ \alpha^2_n \right]$, the variance value of $\alpha_n$ can be exploited as follows:
\begin{align}
    \mathrm{Var}\left[\alpha_n\right] &= \mathrm{E} \left[\alpha^2_n\right] - (\mathrm{E} \left[\alpha_n\right])^2, \\
    \mathrm{E} \left[\alpha^2_n\right] &= \mathrm{Var}\left[\alpha_n\right] +  (\mathrm{E} \left[\alpha_n\right])^2, \\
     &= \frac{\sigma^2_h}{2} \left(2 - \frac{\pi}{2}\right) +\frac{\sigma^2_h \pi}{4}, \\
     &= \sigma^2_h - \frac{\sigma^2_h \pi}{4} + \frac{\sigma^2_h \pi}{4} = \sigma^2_h. 
\end{align}

The same derivations are also valid for $\mathrm{E} \left[\beta^2_{n,k}\right]$, and its value can be obtained as $\sigma^2_{g_{k}}$. Therefore, we can obtain the following result:
\begin{align}
\label{eq:Lemma1_eq10}
    \mathrm{E}\left[\left[\alpha_n \beta_{n,k}\right]^2 \right] = \mathrm{E} \left[ \alpha^2_n \right] \mathrm{E} \left[ \beta^2_{n,k} \right] = \sigma^2_h \sigma^2_{g_k}.
\end{align}

Lastly, by substituting  (\ref{eq:Lemma1_eq4}) and (\ref{eq:Lemma1_eq10}) into (\ref{eq:Lemma1_eq3}), the variance of the multiplication of $\alpha_n$ and $\beta_{n,k}$ can be found as follows:
\begin{align}
    \mathrm{Var} \left[\alpha_n \beta_{n,k}\right]  &= \sigma^2_h \sigma^2_{g_k} - \frac{\sigma^2_{h}\sigma^2_{g_{k}} \pi^2}{16}, \\
    &= \sigma^2_h \sigma^2_{g_k} \left(1-\frac{\pi^2}{16} \right).
\end{align}

When the number of RIS elements allocated to the each UE ($N_g$) is sufficiently large, $\mathcal{A}$ becomes a real Gaussian distributed random variable due to CLT with mean $ \left(N_g \sigma_h \sigma_{g_k} \pi \right)/4$ and variance $N_g \sigma_h \sigma_{g_k} \left(1-\frac{\pi^2}{16} \right)$.

 \section{Proof of Lemma \ref{lemma:OPB}}
	\label{AppOP}

The interference term in the received signal for the $k^{\mathrm{th}}$ UE is given as:
\begin{align}
    \mathcal{B} &=  \sum_{k' \neq k}
 \sum_{n\in \mathbb{I}_{k'}} \alpha_n e^{j \theta_n} \beta_{n,k} e^{j\psi_{n,k}}  e^{j\left( \xi_{k'} - \theta_n - \psi_{n,k'}\right)}, \\
 &=\sum_{k' \neq k}
 \sum_{n\in \mathbb{I}_{k'}} \alpha_n \beta_{n,k} e^{j\psi_{n,k}} e^{j\left( \xi_{k'} - \psi_{n,k'}\right)}, \\
  &=\sum_{k' \neq k}
 \sum_{n\in \mathbb{I}_{k'}} \alpha_n g_{n,k} e^{j\left( \xi_{k'} - \psi_{n,k'}\right)} = \sum_{k' \neq k}
 \sum_{n\in \mathbb{I}_{k'}} \delta_{n,k,k'},
\end{align}
where $g_{n,k} =  \beta_{n,k} e^{j\psi_{n,k}} $ is a random variable which follows $\mathcal{CN}\left(0,\sigma^2_{g_k}\right)$. Since there is no coherent alignment, phase terms in $\mathcal{B}$ are not canceled out and the term $e^{j\left(\xi_{k'} + \psi_{n,k} - \psi_{n,k'}\right)}$ remains. Considering $\mathrm{E} \left[g_{n,k}\right] = 0$, the expected value of $\delta_{n,k,k'}$ is found as
\begin{align}
    \mathrm{E} \left[\delta_{n,k,k'}\right] = \mathrm{E}\left[\alpha_n\right] \mathrm{E}\left[g_{n,k}\right] \mathrm{E}\left[e^{j\left( \xi_{k'} - \psi_{n,k'}\right)}\right] = 0.
\end{align}

Moreover, the variance of $\delta_{n,k,k'}$ is given as:
\begin{align}
    \mathrm{Var} \left[\delta_{n,k,k'}\right] = \: &  \mathrm{E}\left[\left[\alpha_n g_{n,k} e^{j( \xi_{k'} - \psi_{n,k'})}\right]^2 \right] - \nonumber \\ & \left(\mathrm{E}\left[\alpha_n g_{n,k} e^{j\left( \xi_{k'} - \psi_{n,k'}\right)}\right] \right)^2,
    \\ = \: & \mathrm{E} \left[\alpha^2_n\right] \mathrm{E} \left[g^2_{n,k}\right] \mathrm{E} \left[e^{2j( \xi_{k'} - \psi_{n,k'})}\right],\\
    = \: & \sigma^2_h \sigma^2_{g_k}.
\end{align}

Based on the CLT, the summation of a total number of $\left(K-1\right)N_g$ complex terms results in $\mathcal{B}$ converging to a complex Gaussian distribution with zero mean and variance $N_g \left(K-1\right) \sigma^2_h \sigma^2_{g_k}$.

\bibliographystyle{IEEEtran}
\bibliography{IEEEabrv,Ref}

\begin{thebibliography}{10}
\providecommand{\url}[1]{#1}
\csname url@samestyle\endcsname
\providecommand{\newblock}{\relax}
\providecommand{\bibinfo}[2]{#2}
\providecommand{\BIBentrySTDinterwordspacing}{\spaceskip=0pt\relax}
\providecommand{\BIBentryALTinterwordstretchfactor}{4}
\providecommand{\BIBentryALTinterwordspacing}{\spaceskip=\fontdimen2\font plus
\BIBentryALTinterwordstretchfactor\fontdimen3\font minus
  \fontdimen4\font\relax}
\providecommand{\BIBforeignlanguage}[2]{{%
\expandafter\ifx\csname l@#1\endcsname\relax
\typeout{** WARNING: IEEEtran.bst: No hyphenation pattern has been}%
\typeout{** loaded for the language `#1'. Using the pattern for}%
\typeout{** the default language instead.}%
\else
\language=\csname l@#1\endcsname
\fi
#2}}
\providecommand{\BIBdecl}{\relax}
\BIBdecl

\bibitem{saad2019vision}
W.~Saad, M.~Bennis, and M.~Chen, ``A vision of {6G} wireless systems:
  Applications, trends, technologies, and open research problems,'' \emph{IEEE
  Network}, vol.~34, no.~3, pp. 134--142, May 2019.

\bibitem{de2021survey}
C.~De~Alwis, A.~Kalla, Q.-V. Pham, P.~Kumar, K.~Dev, W.-J. Hwang, and
  M.~Liyanage, ``Survey on {6G} frontiers: Trends, applications, requirements,
  technologies and future research,'' \emph{IEEE Open J. Commun. Soc.}, vol.~2,
  pp. 836--886, Apr. 2021.

\bibitem{al2023resource}
M.~Al-Ali and E.~Yaacoub, ``Resource allocation scheme for {eMBB} and {uRLLC}
  coexistence in {6G} networks,'' \emph{Wireless Networks}, pp. 1--20, Apr.
  2023.

\bibitem{dang2020should}
S.~Dang, O.~Amin, B.~Shihada, and M.-S. Alouini, ``What should {6G} be?''
  \emph{Nature Electron.}, vol.~3, no.~1, pp. 20--29, Jan. 2020.

\bibitem{basar2019wireless}
E.~Basar, M.~Di~Renzo, J.~De~Rosny, M.~Debbah, M.-S. Alouini, and R.~Zhang,
  ``Wireless communications through reconfigurable intelligent surfaces,''
  \emph{IEEE Access}, vol.~7, pp. 116\,753--116\,773, Aug. 2019.

\bibitem{elmossallamy2020reconfigurable}
M.~A. ElMossallamy, H.~Zhang, L.~Song, K.~G. Seddik, Z.~Han, and G.~Y. Li,
  ``Reconfigurable intelligent surfaces for wireless communications:
  Principles, challenges, and opportunities,'' \emph{IEEE Trans. Cogn. Commun.
  Netw.}, vol.~6, no.~3, pp. 990--1002, May 2020.

\bibitem{basar2017index}
E.~Basar, M.~Wen, R.~Mesleh, M.~Di~Renzo, Y.~Xiao, and H.~Haas, ``Index
  modulation techniques for next-generation wireless networks,'' \emph{IEEE
  Access}, vol.~5, pp. 16\,693--16\,746, Aug. 2017.

\bibitem{dai2018survey}
L.~Dai, B.~Wang, Z.~Ding, Z.~Wang, S.~Chen, and L.~Hanzo, ``A survey of
  non-orthogonal multiple access for {5G},'' \emph{IEEE commun. surveys \&
  tutorials}, vol.~20, no.~3, pp. 2294--2323, May 2018.

\bibitem{tapio2021survey}
V.~Tapio, I.~Hemadeh, A.~Mourad, A.~Shojaeifard, and M.~Juntti, ``Survey on
  reconfigurable intelligent surfaces below 10 {GH}z,'' \emph{EURASIP J.
  Wireless Commun. Netw.}, vol. 2021, pp. 1--18, Sept. 2021.

\bibitem{vaezi2019non}
M.~Vaezi, R.~Schober, Z.~Ding, and H.~V. Poor, ``Non-orthogonal multiple
  access: Common myths and critical questions,'' \emph{IEEE Wireless Commun.},
  vol.~26, no.~5, pp. 174--180, Sept. 2019.

\bibitem{liu2021sparse}
Z.~Liu and L.-L. Yang, ``Sparse or dense: A comparative study of code-domain
  {NOMA} systems,'' \emph{IEEE Trans. Wireless Commun.}, vol.~20, no.~8, pp.
  4768--4780, Mar. 2021.

\bibitem{ek}
M.~Fu, Y.~Zhou, Y.~Shi, and K.~B. Letaief, ``Reconfigurable intelligent surface
  empowered downlink non-orthogonal multiple access,'' \emph{IEEE Trans. on
  Commun.}, vol.~69, no.~6, pp. 3802--3817, Mar. 2021.

\bibitem{arslan2020index}
E.~Arslan, A.~T. Dogukan, and E.~Basar, ``Index modulation-based flexible
  non-orthogonal multiple access,'' \emph{IEEE Wireless Commun. Lett.}, vol.~9,
  no.~11, pp. 1942--1946, Jul. 2020.

\bibitem{9201413}
S.~Zeng, H.~Zhang, B.~Di, Z.~Han, and L.~Song, ``Reconfigurable intelligent
  surface ({RIS}) assisted wireless coverage extension: {RIS} orientation and
  location optimization,'' \emph{IEEE Commun. Lett.}, vol.~25, no.~1, pp.
  269--273, Jan. 2021.

\bibitem{9309152}
L.~You, J.~Xiong, D.~W.~K. Ng, C.~Yuen, W.~Wang, and X.~Gao, ``Energy
  efficiency and spectral efficiency tradeoff in {RIS}-aided multiuser {MIMO}
  uplink transmission,'' \emph{IEEE Trans. Signal Process.}, vol.~69, pp.
  1407--1421, Dec. 2021.

\bibitem{arslan2022over}
E.~Arslan, I.~Yildirim, F.~Kilinc, and E.~Basar, ``Over-the-air equalization
  with reconfigurable intelligent surfaces,'' \emph{IET Commun.}, vol.~16,
  no.~13, pp. 1486--1497, Jun. 2022.

\bibitem{khaleel2021novel}
A.~Khaleel and E.~Basar, ``A novel {NOMA} solution with {RIS} partitioning,''
  \emph{IEEE J. Sel. Topics Signal Process.}, vol.~16, no.~1, pp. 70--81, Jan.
  2021.

\bibitem{arzykulov2023artificial}
S.~Arzykulov, A.~Celik, G.~Nauryzbayev, and A.~M. Eltawil, ``Artificial noise
  and {RIS}-aided physical layer security: Optimal {RIS} partitioning and power
  control,'' \emph{IEEE Wireless Commun. Lett.}, Mar. 2023.

\bibitem{arslan2023networkindependent}
E.~Arslan, F.~Kilinc, E.~Basar, and H.~Arslan, ``Network-independent and
  user-controlled {RIS}: An experimental perspective,'' in \emph{Proc. 2023
  Int. Symp. Wirel. Pers. Multimedia Commun. (WPMC)}.\hskip 1em plus 0.5em
  minus 0.4em\relax IEEE, Nov. 2023.

\bibitem{kayraklik2022indoor}
S.~Kayrakl{\i}k, I.~Yildirim, Y.~Gevez, E.~Basar, and A.~G{\"o}r{\c{c}}in,
  ``Indoor coverage enhancement for {RIS}-assisted communication systems:
  Practical measurements and efficient grouping,'' in \emph{Proc. 2023 Int.
  Conf. Commun. (ICC)}.\hskip 1em plus 0.5em minus 0.4em\relax IEEE, Feb. 2023.

\bibitem{ozpoyraz2022deep}
B.~Ozpoyraz, A.~T. Dogukan, Y.~Gevez, U.~Altun, and E.~Basar, ``Deep
  learning-aided {6G} wireless networks: A comprehensive survey of
  revolutionary phy architectures,'' \emph{IEEE Open J. Commun. Soc.}, Sept.
  2022.

\bibitem{timotheou2015fairness}
S.~Timotheou and I.~Krikidis, ``Fairness for non-orthogonal multiple access in
  {5G} systems,'' \emph{IEEE Signal Process. Lett.}, vol.~22, no.~10, pp.
  1647--1651, Oct. 2015.

\bibitem{yang2016general}
Z.~Yang, Z.~Ding, P.~Fan, and N.~Al-Dhahir, ``A general power allocation scheme
  to guarantee quality of service in downlink and uplink {NOMA} systems,''
  \emph{IEEE Trans. Wireless Commun.}, vol.~15, no.~11, pp. 7244--7257, Aug.
  2016.

\bibitem{ali2016dynamic}
M.~S. Ali, H.~Tabassum, and E.~Hossain, ``Dynamic user clustering and power
  allocation for uplink and downlink non-orthogonal multiple access ({NOMA})
  systems,'' \emph{IEEE Access}, vol.~4, pp. 6325--6343, Aug. 2016.

\bibitem{ferdi2020error}
K.~Ferdi and K.~Hakan, ``Error probability analysis of non-orthogonal multiple
  access with channel estimation errors,'' in \emph{2020 Proc. IEEE Int. Black
  Sea Conf. Commun. Netw. (BlackSeaCom)}.\hskip 1em plus 0.5em minus
  0.4em\relax IEEE, May 2020, pp. 1--5.

\bibitem{kara2018ber}
F.~Kara and H.~Kaya, ``{BER} performances of downlink and uplink {NOMA} in the
  presence of {SIC} errors over fading channels,'' \emph{IET Commun.}, vol.~12,
  no.~15, pp. 1834--1844, Aug. 2018.

\bibitem{kara2019performance}
------, ``Performance analysis of {SSK}-{NOMA},'' \emph{IEEE Trans. Veh.
  Technol.}, vol.~68, no.~7, pp. 6231--6242, May 2019.

\bibitem{kara2020improved}
------, ``Improved user fairness in decode-forward relaying non-orthogonal
  multiple access schemes with imperfect {SIC} and {CSI},'' \emph{IEEE Access},
  vol.~8, pp. 97\,540--97\,556, May 2020.

\bibitem{arslan2022reconfigurable}
E.~Arslan, F.~Kilinc, S.~Arzykulov, A.~T. Dogukan, A.~Celik, E.~Basar, and
  A.~M. Eltawil, ``Reconfigurable intelligent surface enabled over-the-air
  uplink {NOMA},'' \emph{IEEE Trans. Green Commun. Netw.}, Dec. 2022.

\bibitem{gui2018deep}
G.~Gui, H.~Huang, Y.~Song, and H.~Sari, ``Deep learning for an effective
  nonorthogonal multiple access scheme,'' \emph{IEEE Trans. Veh. Technol.},
  vol.~67, no.~9, pp. 8440--8450, Jun. 2018.

\bibitem{chauhan2022ris}
A.~Chauhan, S.~Ghosh, and A.~Jaiswal, ``{RIS} partition-assisted non-orthogonal
  multiple access ({NOMA}) and quadrature-{NOMA} with imperfect {SIC},''
  \emph{IEEE Trans. Wireless Commun.}, Dec. 2022.

\bibitem{kumar2023ris}
M.~H. Kumar, S.~Sharma, K.~Deka, and M.~K. Sharma, ``{RIS}-assisted user
  pairing {NOMA} system for {THz} communications,'' in \emph{Proc. 2023 Nat.
  Conf. Commun. (NCC)}.\hskip 1em plus 0.5em minus 0.4em\relax IEEE, Feb. 2023,
  pp. 1--6.

\bibitem{mu2020exploiting}
X.~Mu, Y.~Liu, L.~Guo, J.~Lin, and N.~Al-Dhahir, ``Exploiting intelligent
  reflecting surfaces in {NOMA} networks: Joint beamforming optimization,''
  \emph{IEEE Trans. Wireless Commun.}, vol.~19, no.~10, pp. 6884--6898, 2020.

\bibitem{song2021physical}
C.~Song, ``Physical layer security of {RIS}-assisted {NOMA} networks over
  fisher-snedecor composite fading channel,'' in \emph{Proc. 2021 Int. Conf.
  Commun. Comput. Cybersecur. Informatics (CCCI)}.\hskip 1em plus 0.5em minus
  0.4em\relax IEEE, Oct. 2021, pp. 1--6.

\bibitem{gao2023joint}
Q.~Gao, Y.~Liu, X.~Mu, M.~Jia, D.~Li, and L.~Hanzo, ``Joint location and
  beamforming design for {STAR}-{RIS} assisted {NOMA} systems,'' \emph{IEEE
  Trans. Commun.}, vol.~71, no.~4, pp. 2532--2546, Feb. 2023.

\bibitem{bariah2023performance}
L.~Bariah, F.~Boukhalfa, W.~Jaafar, S.~Muhaidat, and H.~Yanikomeroglu, ``On the
  performance of {RIS}-enabled {NOMA} for aerial networks,'' in \emph{Proc.
  2023 IEEE Wireless Commun. Netw. Conf. (WCNC)}.\hskip 1em plus 0.5em minus
  0.4em\relax IEEE, May 2023, pp. 1--6.

\bibitem{li2021effective}
G.~Li, H.~Liu, G.~Huang, X.~Li, B.~Raj, and F.~Kara, ``Effective capacity
  analysis of reconfigurable intelligent surfaces aided {NOMA} network,''
  \emph{EURASIP J. Wireless Commun. Netw.}, vol. 2021, pp. 1--16, Dec. 2021.

\bibitem{aldababsa2023simultaneous}
M.~Aldababsa, A.~Khaleel, and E.~Basar, ``Simultaneous transmitting and
  reflecting reconfigurable intelligent surfaces-empowered {NOMA} networks,''
  \emph{IEEE Syst. J.}, Oct. 2023.

\bibitem{araghi2022reconfigurable}
A.~Araghi, M.~Khalily, M.~Safaei, A.~Bagheri, V.~Singh, F.~Wang, and
  R.~Tafazolli, ``Reconfigurable intelligent surface ({RIS}) in the sub-6 {GH}z
  band: Design, implementation, and real-world demonstration,'' \emph{IEEE
  Access}, vol.~10, pp. 2646--2655, 2022.

\bibitem{basar2019transmission}
E.~Basar, ``Transmission through large intelligent surfaces: A new frontier in
  wireless communications,'' in \emph{2019 European Conference on Networks and
  Communications (EuCNC)}.\hskip 1em plus 0.5em minus 0.4em\relax IEEE, 2019,
  pp. 112--117.

\bibitem{ding2020unveiling}
Z.~Ding, R.~Schober, and H.~V. Poor, ``Unveiling the importance of {SIC} in
  {NOMA} systems—part 1: State of the art and recent findings,'' \emph{IEEE
  Commun. Lett.}, vol.~24, no.~11, pp. 2373--2377, Jul. 2020.

\bibitem{simon2002probability}
M.~K. Simon, \emph{Probability distributions involving Gaussian random
  variables: A handbook for engineers and scientists}.\hskip 1em plus 0.5em
  minus 0.4em\relax Springer, May 2002.

\bibitem{lee1979distribution}
R.-Y. Lee, B.~S. Holland, and J.~A. Flueck, ``Distribution of a ratio of
  correlated gamma random variables,'' \emph{SIAM J. Appl. Math}, vol.~36,
  no.~2, pp. 304--320, Apr. 1979.

\bibitem{3gpp2019nr}
G.~T. 38.104, ``{NR}; base station ({BS}) radio transmission and reception,''
  2019.

\end{thebibliography}
	
\end{document}